\documentclass[final,5p,times]{elsarticle} 

\usepackage{amssymb}
\usepackage{amsthm}
\usepackage{amsmath}
\usepackage{soul}
\usepackage{xcolor, soul}
\usepackage{hyperref}

\journal{  }

\begin{document}

\begin{frontmatter}


\title{A Polynomial Framework for Design of Drag Reducing Periodic Two-dimensional Textured Surfaces}
\author{Shabnam Raayai-Ardakani \corref{cor1}}
\cortext[cor1]{Corresponding author.}
\ead{sraayai@fas.harvard.edu}
\address{{Rowland Institute, Harvard University},
            {100 Edwin H. Land Blvd.}, 
            {Cambridge},
            {02142}, 
            {MA},
            {USA}}

\begin{abstract}

Periodic and symmetric two-dimensional textures with various cross-sectional profiles have been employed to improve and optimize the physical response of the surfaces such as drag force, superhydrophobicity, and adhesion. While the effect of the height and spacing of the textures have been extensively studied, the effect of the shape of the textures has only been considered in qualitative manners. Here, a polynomial framework is proposed to mathematically define the cross-sectional profiles of the textures and offer a quantitative measure for comparing the physical response of the textured surfaces with various cross-sectional profiles. As a case study, textured surfaces designed with this framework are tested for their hydrodynamic frictional response in the cylindrical Couette flow regime in Taylor- Couette flows. With the reduction in torque as the objective, experimental and numerical results confirm that textures with height-to-half-spacing of lower and equal to unity with concave profiles offer a lower torque compared to both smooth surfaces and triangular textures. In addition, across multiple polynomial orders, textures defined by second order polynomials offer a wide range of responses, eliminating the need for considering polynomials of higher orders and complexity. While the case study here is focused on the laminar flow regime and the frictional torque, the same type of analysis can be applied to other surface properties and physical responses as well.

\end{abstract}






\begin{keyword}
Textured Surfaces, Riblets, Taylor-Couette Flows, Drag Reduction



\end{keyword}

\end{frontmatter}


{\color{black}
\section*{Introduction}\label{Introduction}

Textures are well documented for their ability to alter and regulate physical response of surfaces, both in the natural and engineering settings; Drag reducing effect of ribs on the denticles of fast swimming shark species \cite{bechert_drag_1985, wen_biomimetic_2014, dinkelacker1988possibility} as well as bio-inspired riblet-covered surfaces \cite{goldstein1995direct, choi1993direct, Walsh1980, walsh1983riblets, walsh1984optimization, raayai2020geometry, raayai2018geometry, bacher1986turbulent, djenidi1991high, djenidi1994laminar, djenidi1989numerical, Chu_Karniadakis_1993,zheng2020effect, rastegari2019drag, daniello2009drag}, super-hydrophobicity, super-repellency, and self cleaning of \textit{Colocasia esculenta}, \textit{Euphorbia myrsinites}, \textit{Nelumbo nucifera} (Lotus), and \textit{Salvinia} leaves, packed with convex papillose cells which are covered by three-dimensional wax crystal \cite{barthlott_plant_2017,teisala_hierarchical_2019,ensikat_superhydrophobicity_2011,xiang_superrepellency_2020, zhu_antisoiling_2020,erramilli2019influence}, slippery and insect trapping mechanism of the concave textures on the rim (peristome) of \textit{Nepenthes} pitcher plants \cite{bohn_insect_2004,gorb_structure_2004,gorb_structure_nodate, bauer_insect-trapping_2009}, which can also act as superior fog harvesters  \cite{li2020liquid},  as well as excellent adhesion of textures on the feet of gecko and some artopods \cite{gilman2015geckos, imburgia2019effects,erramilli2019influence} and the textured gecko-inspired engineered adhesives \cite{king2014creating, suresh2021forcing, croll2019switchable, arzt2021functional, abbott2007mass} are but a few examples. 

Textures come in random and ordered configurations, and the characteristics of the profiles and patterns are described with a variety of naming conventions. Terms such as roughness, waviness, and lay \mbox{\cite{degarmo2003materials}} in manufacturing, dimples and seams in sports \cite{mehta1985aerodynamics, hong_aerodynamic_2017}, or wrinkles, dimples, and creases in patterns formed with soft materials \cite{genzer2006soft, chan2006spontaneous, raayai2016mechanics} are a few of the words commonly used to describe the corresponding topographies. In dealing with random roughness under flow conditions, additional characteristics, such as the roughness topography, local peak-to-trough heights, max peak-to-trough height, or sand grain roughness measures are also considered \cite{chung2021predicting}.

One group of idealized and ordered engineering textures are two-dimensional (2D) textures; these are simplified models of natural rib-like textures (shark skin, butterfly wings, rice leaves), and have been widely considered for the purposes of drag reduction \cite{BECHERT_On_1985} and anisotropic surface responses \cite{nishimoto2013bioinspired, bixler2014anti}. Such 2D ordered textures are comprised of a unit element of texture that is symmetric and repeated in a periodic pattern (Fig. \ref{fig:peak_trough_schematic}(a)) and unit elements ranging from triangular (also known as V-groove or saw-tooth) \cite{walsh1984optimization, goldstein1995direct, Choi_1992, choi1993direct, djenidi1989numerical, djenidi1991high}, semi-circular \cite{Bechert_Hoppe_Hoeven_Makris_1992, lee2001flow, walsh1983riblets}, rectangular \cite{walsh1979drag, Launder_Li_1989,wang1993drag, el2007drag, Bechert_Bruse_1997}, families of sine waves \cite{raayai2017drag}, and trapezoidal \cite{lazosturbulent1987, Bechert2000, Bechert_Bruse_1997} have been considered before. Due to the difference in the shapes of the textures, various physical dimensions (height, spacing, radius of curvature, etc.) have been used to compare the frictional drag response of the textures; height and spacing of the texture elements are universally applicable to any periodic 2D texture, but other geometric features (corner angles, the radii of curvatures, etc.) are shape-specific and not expandable to other shapes, and descriptive names only offer the means for qualitative comparison of the performance among the variety of textures.

\begin{figure*}[!ht]
     \centering
    \includegraphics[width = 1\textwidth]{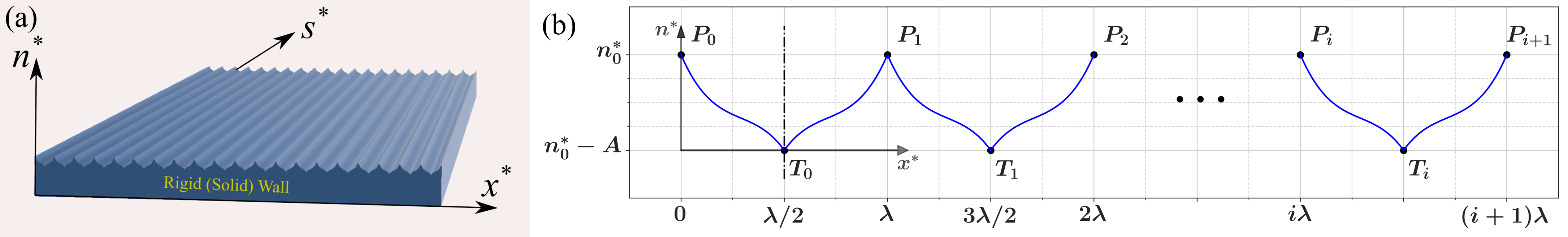}
    \caption{(a) Three dimensional schematic of a periodic, symmetric, 2D texture on a rigid solid wall; (b) Projection of the 2D profile on the ($x^*$-$n^*)$ plane. The horizontal distance between consecutive peaks ($P_i$ and $P_{i+1}$) and troughs ($T_i$ and $T_{i+1}$) is the spacing, $\lambda$, and the vertical distance between peaks and troughs is $A$. The profile is $\lambda$-periodic, and symmetric about lines of $x^*=j\lambda/2$, where $j \in \mathbb{N}$. One example of a symmetry line is presented with a dashed-dotted line at $x^*=\lambda/2$.}
    \label{fig:peak_trough_schematic}
\end{figure*}

Here, a polynomial framework is proposed to quantitatively define the shape of the textures, and allow for an objective comparison among the various families of grooves. While the scope of this work is focused on textures in hydrodynamic applications to optimize the frictional drag force, the mathematical classification of the textures can be expanded to other scientific applications as well. Here, the performance of the surfaces is evaluated using experimental and numerical analysis of the textures in a Taylor-Couette (TC) flow \cite{grossmann2016high, raayai2020geometry, raayai2018geometry} especially focusing on the frictional torque (equivalent to drag in cylindrical coordinate system) response of the surfaces in cylindrical Couette flow regime (CCF) as the key measure of comparison. 

\section*{Methods}\label{sec11}

\subsection*{Experimental procedure} The experiments were conducted using a custom-designed and built Taylor-Couette (TC) cell (Fig. \ref{fig:exp_setup}) attached to a stress-controlled rheometer (TA Discovery Hybrid Rheometer 2). The radius ratio of the TC cell is kept at $\eta = R_{\rm i}/R_{\rm o } = 0.65$ and the height ratio $\Gamma = L/(R_{\rm o}-R_{\rm i}) = 5.1$ where $R_{\rm o}$ and $R_{\rm i}$ are the radii of the outer cylinder and the inner cylinder respectively, and $L$ is the length of the rotor (Fig. \ref{fig:exp_setup}). Here, the outer cylinder is kept stationary ($\Omega_{\rm o} = 0$), the inner cylinder rotates at a rotation rate of $\Omega_{\rm i}$, and the torque experienced by the rotor, $\cal{T}_{\rm total}$, is measured via the sensors in the rheometer. Further details of the design of a similar TC cell can be found in Ref. \cite{raayai2018geometry}. 

Textured covered rotors were 3D printed via stereo-lithography (Formlabs Form3 3D printer and photopolymer resin). The design of the full rotors comprise of the inner rotor (which is in contact with the working fluid during the experiments) and a sleeve with a tapped hole that allows for the rotors to be installed on the rheometer and be connected to the sensors (Fig. \ref{fig:exp_setup}). The outermost radius of the textured cylinders (i.e. texture peak) is at $R_{\rm i}$ and the smallest radius of the textures is located at the trough at $R_{\rm i} - A$, where $A$ is the height of the textures (Fig. \ref{fig:peak_trough_schematic}(b)). 

The effect of the gap at the closed bottom of the rotor is subtracted from the measured torque using the procedure outlined in Refs. \cite{raayai2018geometry, raayai2020geometry}; in summary, the distance between the bottom of the rotor and the TC cell ($h_{\rm b}$) is calibrated using

\begin{equation}
    {\cal T}_{\rm bottom} = \mu \dfrac{\pi R_{\rm i}^4 }{2 h_{\rm b}} \Omega_{\rm i}
    \label{bottom_gap}
\end{equation}

\noindent and set to $h_{\rm b} = 340\ \mu {\rm m}$, and using Eq. \eqref{bottom_gap} the excess torque from this bottom effect is then subtracted from all measurements (i.e. ${\cal T} = {\cal T}_{\rm total} - {\cal T}_{\rm bottom}$). 

\begin{figure}
    \centering
    \includegraphics[width = 0.45\textwidth]{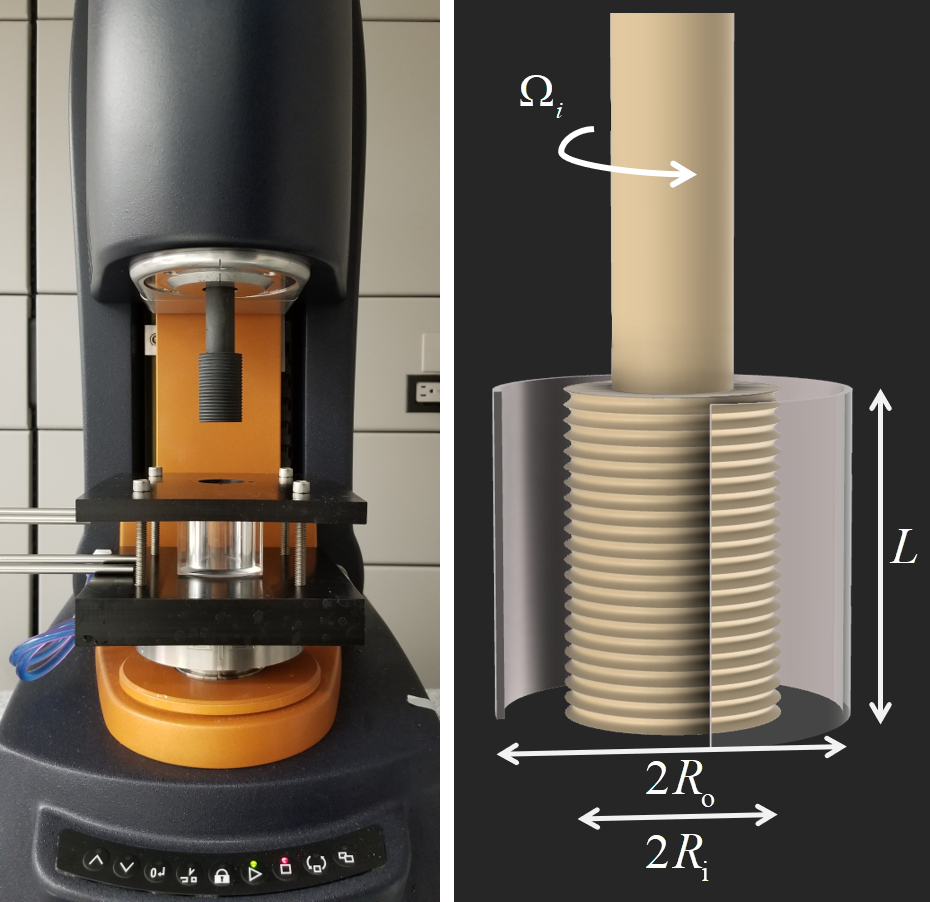}
    \caption{Image of the experimental Taylor-Couette setup, attached to a stress-controlled rheometer on the left, and a schematic of the setup, on the right.}
    \label{fig:exp_setup}
\end{figure}

The working fluid is a mixture of 70\% Glycerol (Sigma Aldrich BioXtra $\geqslant 99 \%$) and 30\% de-ionized water (volumetric ratio) and the viscosity of the working fluid is measured at the beginning and the end of the experimental procedure as a function of temperature. To account for any affect of temperature on the viscosity of the working fluid, throughout the experiment temperature of the fluid and the surrounding environment is measured in 60 second intervals using a two channel thermometer (Digi-sense Tracable Memory-loc Datalogging Thermometer) and the fluid temperature is used to adjust the viscosity of the working fluid during data analysis. The maximum temperature change recorded in either channel of the thermometer, during each experimental procedure is less than 0.5-0.8$^{\circ} \ {\rm C}$ (viscosity variation within $2\%$ of the average viscosity), and while this variation is accounted for in the analysis, turns out to have a negligible effect on the results. (This can also be shown using the Nahme-Griffith number \cite{raayai2020geometry}.) The density of the mixture is calculated using the formula given in Ref. \cite{volk2018density}.

The rotation rates, and torque, as well as fluid properties are used to calculate the Reynolds number and the dimensionless torque (following the definition from references \cite{lathrop1992transition, lathrop1992turbulent}) using  

\begin{equation}
    {\rm Re}_d = \dfrac{\rho  (R_{\rm o} - R_{\rm i}) R_{\rm i} \Omega_{\rm i}}{\mu}
\end{equation}

\noindent and

\begin{equation}
    G = \dfrac{\cal{T}}{4 \pi \rho {\nu}^2 L}. 
\end{equation}

\noindent where $\rho$, and $\mu$ are the fluid density and dynamic viscosity respectively, and $\nu = \mu/\rho$.

For the purpose of comparison, a smooth rotor with inner radius of $R_{\rm i}$, and length $L$ is also tested with the same procedure. The measured dimensionless torque for the smooth rotor $G_0$ (after subtraction of the bottom effect) is on average within $0.6\%$ of the theoretical torque 

\begin{equation}
    G_{\rm th} = \dfrac{\eta}{(1-\eta^2)(1-\eta)} {\rm Re}_d
\end{equation}

\noindent in CCF regime within $10 \leqslant {\rm Re}_d \leqslant 40$ and within $2\%$ of the theoretical torque within $40 < {\rm Re}_d \leqslant 60$ (average). The measured $G_0$ is used throughout the text to normalize the torque of the rest of the samples. These measured torques and Reynolds numbers were used to identify the onset of transition to the Taylor vortex flow, ${\rm Re}_{d,{\rm tr}} = 64$, following the method described in Refs. \cite{lathrop1992transition, lathrop1992turbulent}.

The experimental procedure used here is based on a set of 60-step ``peak hold'' procedures were at each step a constant rotation rate is imposed on the inner rotor for a span of 15 seconds and torque is read five times during these 15 seconds and an average of these five readings is used for the analysis ($\cal{T_{\rm total}}$). This ``peak hold'' is repeated for velocities corresponding to CCF regime (i.e. Reynolds numbers less than ${\rm Re}_{d,{\rm tr}} = 64$). Measurements for the textured surfaces didn't present any sign of transition happening earlier than ${\rm Re}_d = 64$. Each set of experiment is then repeated 6 times and the results are presented here as mean $\pm$ one standard deviation. 

\subsection*{Numerical simulations} 

For comparison, direct numerical simulation of the CCF over textured surfaces are conducted using the finite volume package, OpenFOAM. The simulations are performed for one single unit of texture, in axisymmetric, steady-state conditions, using the SIMPLE algorithm, with periodic boundary conditions at $x^*=0$, and $x^* = \lambda$. Here the diffusion terms are discretized with a
second-order central difference, and the convective terms are discretized with a second-order central difference with correction for the non-orthogonality of the mesh faces due to the textures. 

\section*{Results}

\subsection*{Periodic, symmetric 2D textures}

A 2D periodic texture (Fig. \ref{fig:peak_trough_schematic}(a)), consists of symmetric unit elements (between a peak-trough-peak group, $P_i$, $T_i$, and $P_{i+1}$ - Fig. \ref{fig:peak_trough_schematic}(b)) where the streamwise and spanwise directions are denoted by $s^*$ and $x^*$ and the normal to the two directions is denoted by $n^*$ (all starred variables are dimensioned). The unit of texture has a spacing of $\lambda$ (i.e. $(x^*_{_{P_{i+1}}}-x^*_{_{P_i}}) = (x^*_{_{T_{i+1}}}-x^*_{_{T_i}}) =\lambda$) and height of $A$ (i.e. $(n^*_{_{P_i}}-n^*_{_{T_i}}) = A$), and half of the symmetric profile of the wall in between $P_{\rm i}$ and $T_{\rm i}$ is defined by a function $n^{*}_{\rm w} = f^*(x^*)$. Due to the symmetry, all directions are non-dimensionalized by half of the wavelength ($s = s^*/(\lambda/2)$, $x = x^*/(\lambda/2)$, $n = n^*/(\lambda/2)$), and the height-to-half-spacing ratio is denoted by ${\cal{R}} = A/(\lambda/2)$ (i.e. $(x_{_{T_{i}}}-x_{_{P_i}}) = 1$, and $(n_{_{P_i}}-n_{_{T_i}}) = \cal R$). Here, it is assumed that all peaks and troughs are located at $n=n_0$ and $n=n_0-{\cal R}$ respectively (where $n_0$ is an arbitrary point of reference and can be shifted to any choice of reference point and coordinate system) and for all the cases considered here, the domain located above the texture profile ($n>n_{\rm w}$) is filled with fluid and the domain below this boundary ($n<n_{\rm w}$) is a rigid wall.

With this setting, for textures with the same peak and trough locations, function $f$ is the key to distinguishing between textures of different shapes. For example, in the simplest form, one can set $f$ to be a line connecting $P_i$ and $T_i$, in the form of

\begin{equation}
n_{\rm w} = -{\cal{R}} x + n_0 \ \ \ \ {\rm for} \ \ \ \ 0 \leqslant x \leqslant 1
\label{order_1}
\end{equation}

\noindent where ${\cal R}$, or the height-to-half-spacing ratio is also the magnitude of the slope of the function $f$ (in $0<x<1$). This functional form returns the triangular or V-groove textures that have been considered extensively in literature \cite{walsh1984optimization, goldstein1995direct, Choi_1992, choi1993direct, djenidi1989numerical, djenidi1991high}. In a similar manner, $n_{\rm w} = n_0$ returns the case of a smooth reference surface. 

\subsection*{Concavity and convexity}

\begin{figure*}[!h]
    \centering
    \includegraphics[width = 1 \textwidth]{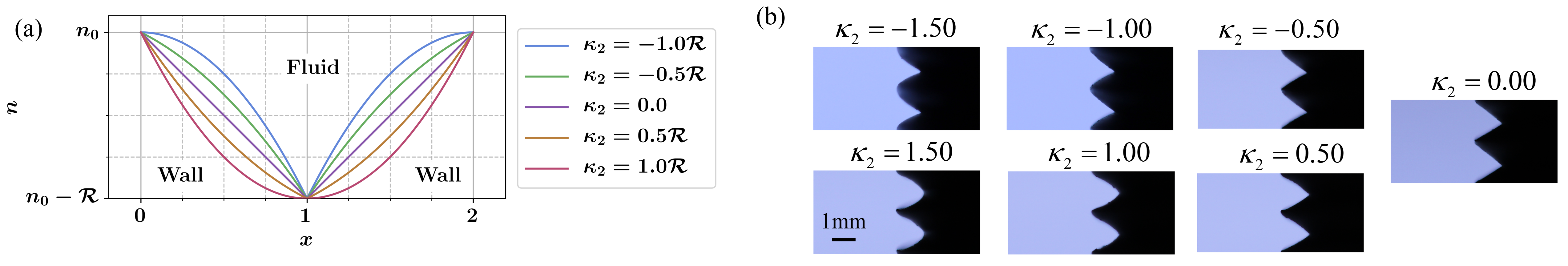}
    \caption{(a) Family of curved textures, defined by second order polynomials with height-to-half-spacing of $\cal R$, and $-{\cal R} \leqslant \kappa_2 \leqslant {\cal R}$. (b) Profiles of two unit textures on the surface of 3D printed rotors covered with second order textures with ${\cal R} = 1.50$ and $-1.50 \leqslant \kappa_2 \leqslant 1.50$.}
    \label{fig:setup}
\end{figure*}

\begin{figure*}[!ht]
    \centering
    \includegraphics[width = 1\textwidth]{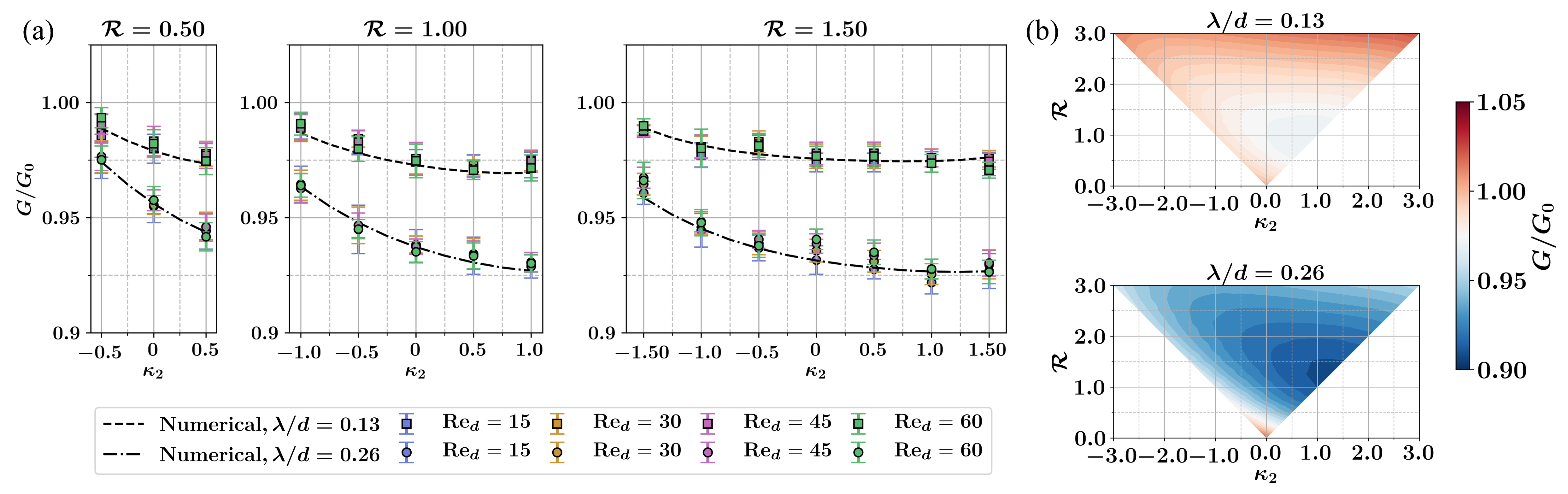}
    \caption{(a) Dimensionless torque experienced by textured rotors with profiles defined by second order polynomials, with ${\cal R} = 0.50$, $1.00$, and $1.50$, and $\lambda/d = 0.13$, and $0.26$ as a function of $\kappa_2$. Torque is normalized by the torque experienced by the smooth rotor. (b) Contours of dimensionless torque of textured rotors with a profile defined by second order polynomials as a function of $\cal R$, and $\kappa_2$, normalized by the torque of a smooth rotor, $G_0$, calculated numerically, for grooves with $\lambda/d = 0.13$ (top), and $\lambda/d = 0.26$ (bottom). Sample lines from the contour plots of $G/G_0$ for select height-to-half-spacings are plotted in Fig. \ref{Line_plots_of_torque} for additional clarity.}
    \label{2nd_profile}
\end{figure*}

\begin{figure*}[!ht]
\centering
\includegraphics[width=0.8\textwidth]{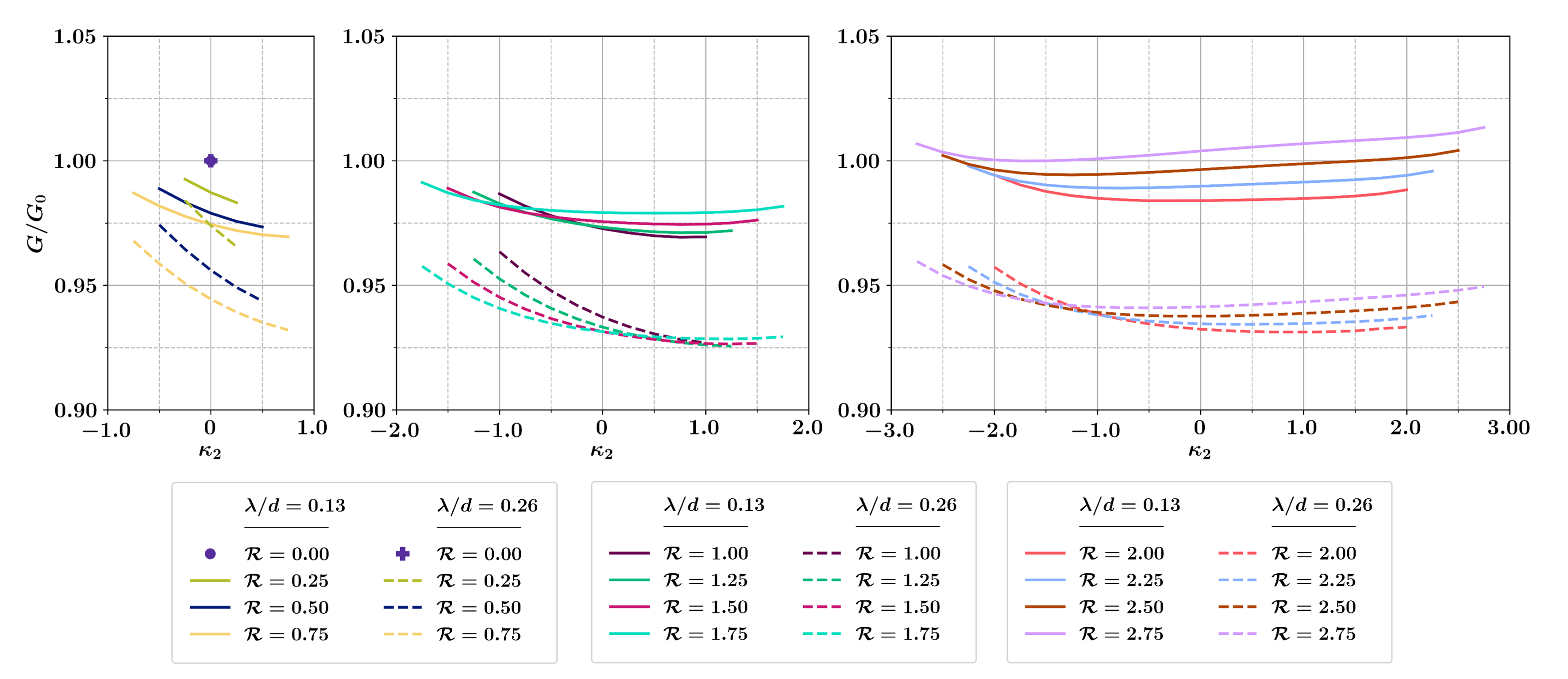}
\caption{Line plots of torque normalized by the torque of a smooth rotor for select height-to-half-spacing ratios as a function of $\kappa_2$, and $\lambda/d = 0.13$, and $0.26$, corresponding to the contour plots in Fig. \ref{2nd_profile}(b).}
\label{Line_plots_of_torque}
\end{figure*}

When considering a curved (or non-linear) profile, $f$, for each unit element of the texture, after an average slope, $\cal R$, (i.e. height-to-half-space ratio), concavity becomes a second geometric feature that is considered. With concavity defined as $d^2 n_{\rm w}/d x^2>0$ (and vice versa convexity as $d^2 n_{\rm w}/d x^2<0$) the simplest function representing a curved texture has a constant second derivative, resulting in a second order polynomial of the form $n_{\rm w} = a_2 x^2 + a_1 x + a_0$ where $2a_2$ is the constant second derivative and $a_1$ and $a_0$ are found based on $\cal R$, and the choice of $n_0$. If the constant second derivative is given as $2\kappa_2$, then $f$ is written as

\begin{equation}
    n_{\rm w} = \kappa_2 x^2 + (- \kappa_2-{\cal{R}}) x + n_0   \ \ \ \ \ \ \ 0\leqslant x \leqslant 1.  \label{order_2}
\end{equation}

\begin{figure*}[!ht]
    \centering
    \includegraphics[width = 1 \textwidth]{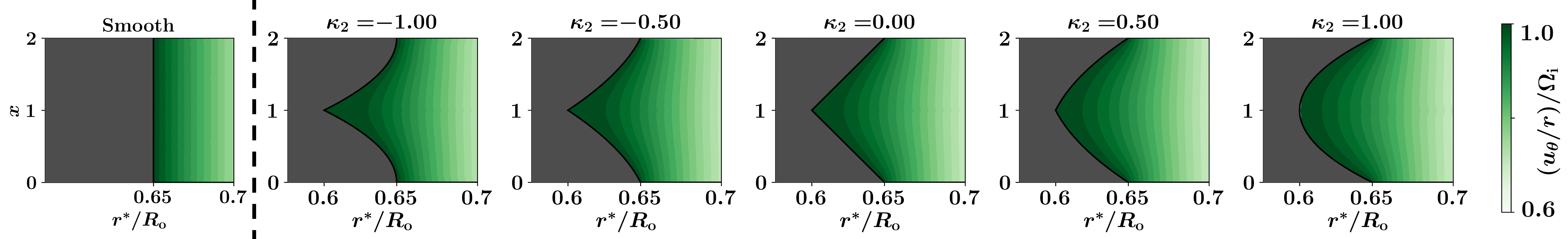}
    \caption{Distribution of the angular velocity ($u_{\theta}/r^*$) normalized by the angular velocity of the inner rotor ($\Omega_{\rm i}$) for textures with ${\cal R} = 1.00$, and $-1\leqslant \kappa_2 \leqslant 1$. Angular velocity of a smooth case is presented on the left for comparison.}
    \label{v_theta_mesh}
\end{figure*}

For textures where the peaks and troughs are the only global (and local) extrema, the profile of the textures, $f$, needs to be locally monotonic between each peak-trough segment and this results in $\kappa_2$ being bounded by the limits $-{\cal{R}} \leqslant \kappa_2 \leqslant {\cal{R}}$ (Fig. \ref{fig:setup}(a)). The definition of the profile as written in Eq. \eqref{order_2} decomposes the effect of the texture geometry into two terms identified by the height-to-half-spacing ratio, $\cal R$, and $\kappa_2$, as a curvature parameter. Therefore, if $k_2 = 0$ (i.e. $d^2 n_{\rm w}/dx^2 = 0 $), $f$ returns the profile of a triangular texture (Eq. \eqref{order_1}). At a constant $\cal R$, varying $\kappa_2$ from its lowest ($\kappa_2 = -{\cal R}$) to highest ($\kappa_2 = {\cal R}$) limits, results in a series of textures spanning from convex ($\kappa_2<0$) to concave ($\kappa_2>0$) profiles (Fig. \ref{fig:setup}).

Experimental and numerical evidence show that varying $\kappa_2$ directly affects the frictional response of the flow in the laminar regime. 3D printed texture-covered inner rotors with ${\cal R} = 0.50$, $1.00$, and $1.50$ and various values of $\kappa_2$ (Fig. \ref{fig:setup}(b)) are tested in the CCF regime and the frictional torque experienced by the rotors ($\cal T$) is measured as a function of the rotation rate, $\Omega_{\rm i}$, of the inner rotors. In this cylindrical setting, $r$, $\theta$, and $x$ are the the respective normal, streamwise, and spanwise directions. The measurements (Fig. \ref{2nd_profile}(a)) are presented in terms of dimensionless torque, $G$, normalized by reference $G_0$ of a smooth rotor, as a function of $\cal R$, $\kappa_2$, and the Reynolds number of the flow, ${\rm Re}_d$.

While independent of the Reynolds number (similar to the previous reports for CCF \cite{raayai2020geometry,raayai2018geometry}), for all samples (Fig. \ref{2nd_profile}(a)), decreasing $\kappa_2$ from zero (i.e. moving from a triangular profile toward a convex profile) results in an increase in the torque compared to that of the V-grooves ($\kappa_2 = 0$). However, for ${\cal R} = 0.50$ and $1.00$ increasing $\kappa_2$ from zero (moving from triangular to a concave texture) results in a reduction in the measured torque with respect to the triangular texture. This is while for ${\cal R} = 1.50$, the same increase in $\kappa_2$ does not offer a substantial change. These trends are similar to two aspects of previous reports of tests using curved textures in turbulent flow in wind tunnel; First, between triangular, and semi-circular textures in concave and convex format, concave textures reduce the drag force the most; triangular textures present less reduction compared to the concave ones, and convex ones do not show any difference compared with a smooth wall \cite{walsh1984optimization}. Second, increasing the radius of curvature of troughs of grooves results in enhancements in the total drag reduction achieved, while increasing the radius of the curvature at the peaks, results in an increase in the total measured drag force \cite{walsh1983riblets}. On the other hand, the current measurements also confirm that (similar to triangular textures \cite{raayai2020geometry, raayai2018geometry}) employing a larger wavelength results in a lower level of torque on the rotors and the introduction of $\kappa_2$ does not affect this trend. 

Changes in torque as a function of $\kappa_2$ are dependent on $\cal R$ of the grooves (as presented in the contours of Fig. \ref{2nd_profile}(b) and corresponding line plots in Fig. \ref{Line_plots_of_torque}); while confirming the trends presented in Fig. \ref{2nd_profile}(a), direct numerical simulations of textures with $0 \leqslant {\cal{R}} \leqslant 3$ and $\lambda/d = 0.13$ and $\lambda/d = 0.26$ show that for profiles with ${\cal{R}} \leqslant 1.00$, going from a convex to a concave profile offers a decrease in the measured torque; textured rotors with $1.00<{\cal R} \leqslant 2.00$ (similar to ${\cal R} = 1.50$ in Fig. \ref{2nd_profile}(a)), do not experience a substantial change in the torque for $0 \leqslant \kappa_2 \leqslant {{\cal R}}$ while moving toward a convex profile results in an increase in the torque (compared with the V-grooves); and lastly for cases with ${\cal R}>2.00$, the trend slowly reverses where convex profiles experience frictional torques lower than the case of a triangular groove and all concave profiles increase the torque.

Variation in $G$ as a function of $\kappa_2$ can by assessed by examining the distribution of angular velocity and consequently the wall shear stress at the textured wall. Contours of the angular velocity of the flow, $u_{\theta}/r$, (Fig. \ref{v_theta_mesh}) show that in comparison to a smooth rotor, the iso-lines inside the grooves are curved, mimicking a pattern dictated by the shape of the texture at the boundary (but not exactly the same shape) and as one moves further away from the boundary (Fig. \ref{v_theta_mesh} at $r^*/R_{\rm o} \gtrapprox 0.68$) the iso-lines return to being parallel to the $x$ direction. The effect of $\kappa_2$ is especially visible at peaks and troughs where for all the textures, the iso-lines are squeezed close to each other at the peaks, but at the troughs the lines are pushed further apart. 

In laminar flows, such as the case of the CCF here, viscous diffusion plays the key role in the development of the flow. With the diffusion occurring in the wall-normal direction and the wall shear stress being proportional to the velocity gradient in the wall-normal direction, employing a coordinate transformation from the generic ($x-n$) plane to a curvilinear, orthogonal coordinate system of ($\eta-\xi$) where the iso-$\xi$ and iso-$\eta$ lines are respectively parallel and perpendicular to the textured boundary, offers additional insight.

\begin{figure}[!ht]
\centering
\includegraphics[width=0.4 \textwidth]{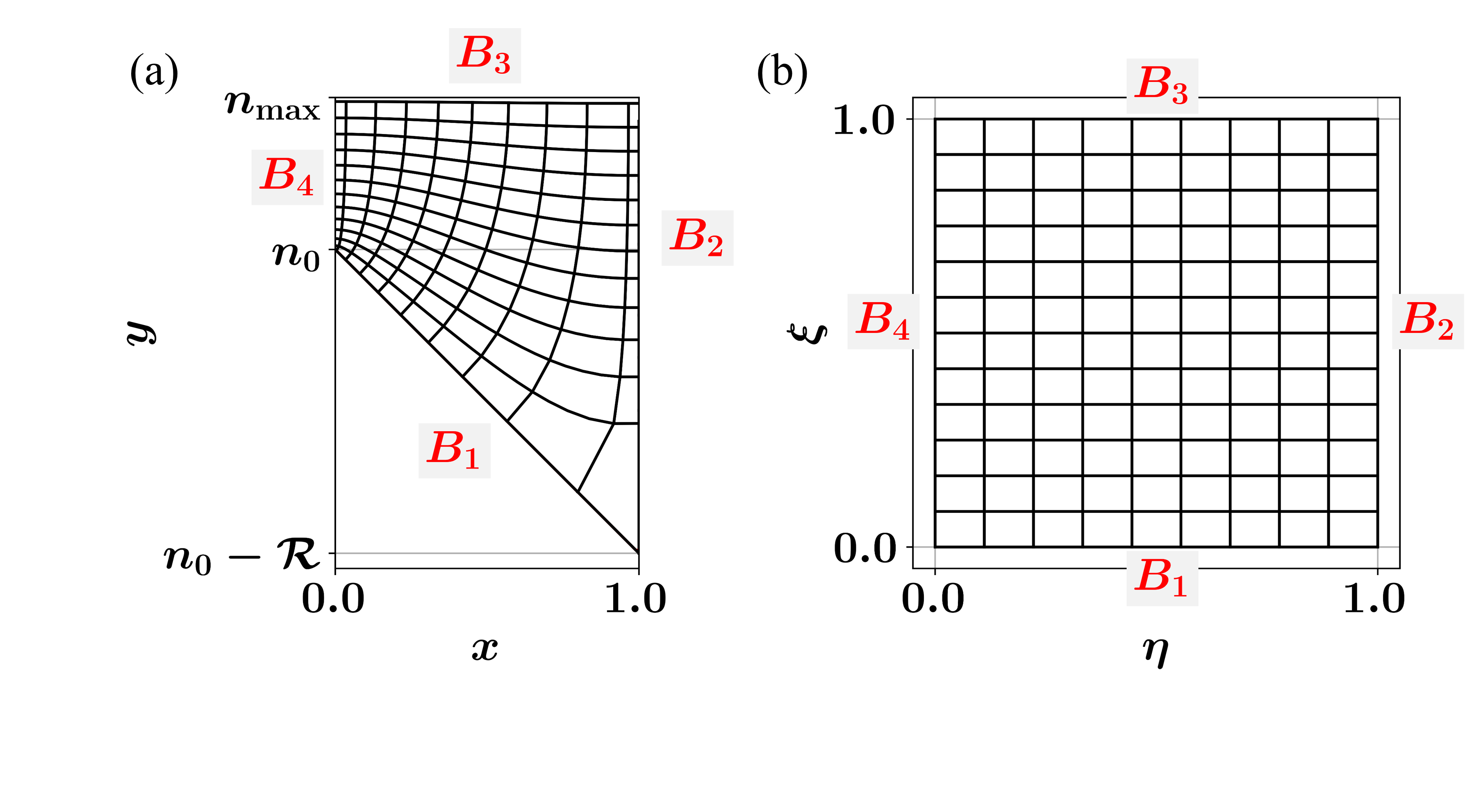}
\caption{Boundaries of the domain mapped between (a) ($x-n$) plane and (b) ($\eta-\xi$) plane resulting in the curvilinear orthogonal coordinate system for an example triangular texture.  }
\label{BC}
\end{figure}

\begin{figure*}[!h]
    \centering
    \includegraphics[width = 1.0 \textwidth]{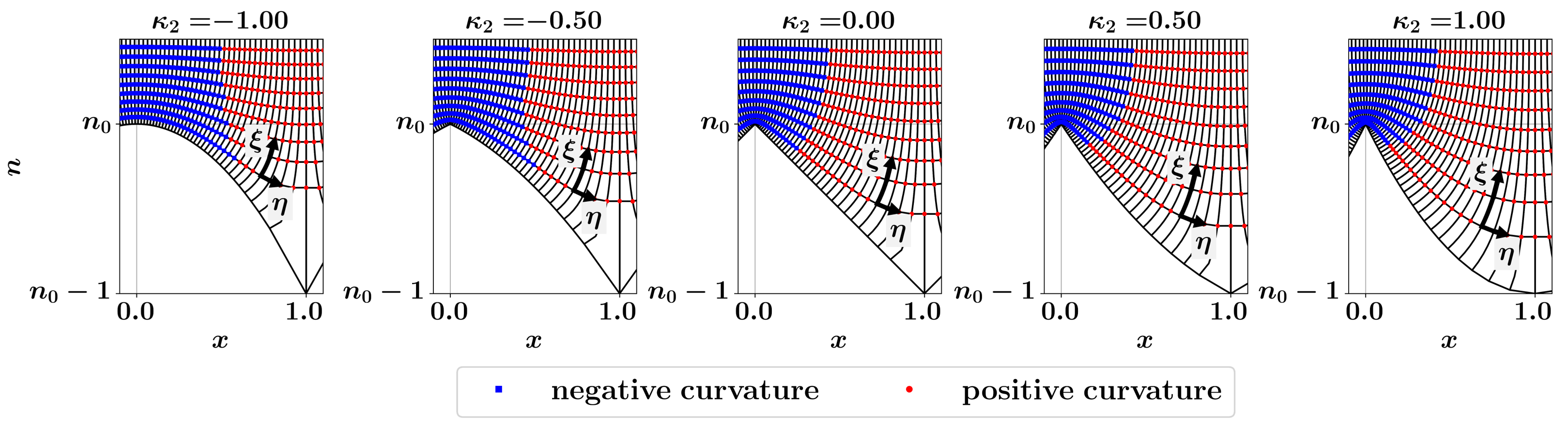}
    \caption{Iso-lines of curvi-linear, orthogonal coordinate system ($\eta-\xi$), perpendicular to the texture boundary for textures with ${\cal R} = 1.00$, and $-1.00\leqslant \kappa_2 \leqslant 1.00$. Locations along the iso-$\xi$ with negative and positive curvature have been marked with blue squares and red circles respectively.}
    \label{meshes}
\end{figure*}

To find the curvilinear-orthogonal coordinate system corresponding to each of the texture domains, inverse of an elliptical mapping between the $(x-n)$ coordinate and the new $(\eta - \xi)$ coordinates is employed (Fig. \ref{BC}). The forward map is defined by Laplace equations of the form 

\begin{equation}
    \nabla^2 \eta(x,n) = 0
    \label{eta}
\end{equation}

\begin{equation}
    \nabla^2 \xi(x,n) = 0
    \label{ksi}
\end{equation}

\noindent with boundary conditions of

\begin{equation}
    \dfrac{\partial \eta}{ \partial h}\bigg\rvert_{B_1} = 0  \ \ \ \ \& \ \ \ \ \ \xi \bigg\rvert_{B_1} = 0
    \label{BC1}
\end{equation}

\begin{equation}
    \dfrac{\partial \xi}{ \partial h}\bigg\rvert_{B_2} = 0  \ \ \ \ \& \ \ \ \ \ \eta\bigg\rvert_{B_2} = 1
    \label{BC2}
\end{equation}

\begin{equation}
    \dfrac{\partial \eta}{ \partial h}\bigg\rvert_{B_3} = 0  \ \ \ \ \& \ \ \ \ \ \xi \bigg\rvert_{B_3} = 1
    \label{BC3}
\end{equation}

\begin{equation}
    \dfrac{\partial \xi}{ \partial h}\bigg\rvert_{B_4} = 0  \ \ \ \ \& \ \ \ \ \ \eta\bigg\rvert_{B_4} = 0
    \label{BC4}
\end{equation}

\noindent where $B_1$ is the textured boundary and the rest of the boundaries are numbered $B_2$-$B_4$ in the counter clockwise direction (Fig. \ref{BC}), and $h$ is the local normal to the surface at each boundary. Thus the inverse of this mapping \cite{thompson1974automatic} is calculated by solving  Eq. \eqref{coord_trans_1}, and Eq. \eqref{coord_trans_2}

\begin{equation}
    g_{_{22}} \dfrac{\partial^2 x}{\partial \eta^2} - 2 g_{_{12}} \dfrac{\partial^2 x}{\partial \eta \partial \xi} + g_{_{11}} \dfrac{\partial^2 x}{\partial \xi^2} = 0 
    \label{coord_trans_1}
\end{equation}

\begin{equation}
    g_{_{22}} \dfrac{\partial^2 n}{\partial \eta^2} - 2 g_{_{12}} \dfrac{\partial^2 n}{\partial \eta \partial \xi} + g_{_{11}} \dfrac{\partial^2 n}{\partial \xi^2} = 0 
    \label{coord_trans_2}
\end{equation}

\noindent where $g_{_{11}}$, $g_{_{12}}$, and $g_{_{22}}$ are the metric tensors connecting the two coordinate systems 

\begin{equation}
    g_{_{11}}= \left( \dfrac{\partial x}{\partial \eta} \right) ^2  + \left( \dfrac{\partial n}{\partial \eta} \right) ^2
\end{equation}

\begin{equation}
    g_{_{22}}= \left( \dfrac{\partial x}{\partial \xi} \right) ^2  + \left( \dfrac{\partial n}{\partial \xi} \right) ^2
\end{equation}

\begin{equation}
    g_{_{12}}= \left( \dfrac{\partial x}{\partial \eta} \right) \left( \dfrac{\partial x}{\partial \xi} \right)  + \left( \dfrac{\partial n}{\partial \eta} \right) \left( \dfrac{\partial n}{\partial \xi} \right).
\end{equation}

The equations are solved iteratively with a python script, using an algorithm similar to the approach presented in \cite{MeshGenerate} subject to the boundary conditions in Eq. \eqref{BC1}, Eq. \eqref{BC2}, Eq. \eqref{BC3}, and Eq. \eqref{BC4}. 

The trend in the iso-$\xi$ lines match the iso-velocity lines (in Fig. \ref{v_theta_mesh}) very well and close to the peaks of all the textures ($x \approx 0$ in Figs. \ref{v_theta_mesh} and \ref{meshes}), both iso-$\eta$ and iso-$\xi$ lines are squeezed close to each other, while in the trough area ($x \approx 1$ in Figs. \ref{v_theta_mesh} and \ref{meshes}), the distances between the iso-lines are more relaxed.

\begin{figure*}[!ht]
    \centering
    \includegraphics[width = 1 \textwidth]{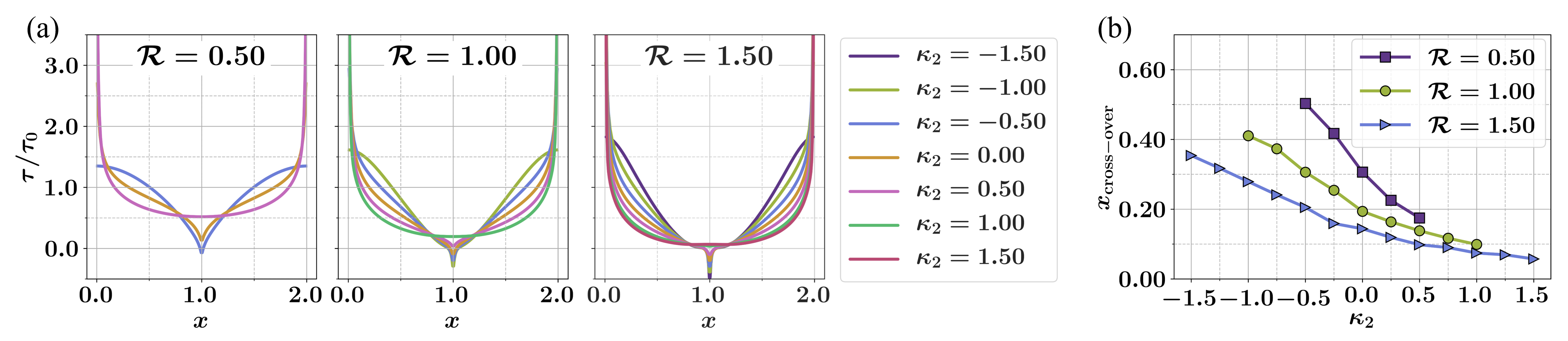}
    \caption{(a) Distribution of wall shear stress as a function of the spanwise location, $x$, within each texture unit, for textures with second order profiles, ${\cal R} = 0.50$, $1.00$, and $1.50$, and $\lambda/d=0.26$, as a function of $-{\cal R}\leqslant \kappa_2 \leqslant {\cal R}$. (b) Location of point of cross-over of the shear stress from $\tau>\tau_{_0}$ to $\tau<\tau_{_0}$, as a function of $\kappa_2$ for textures with ${\cal R} = 0.50$, $1.00$, and $1.50$, and $\lambda/d = 0.26$ within $0 \leqslant x \leqslant 1$.}
    \label{WSS_second_order}
\end{figure*}

\begin{figure*}[!ht]
\centering
\includegraphics[width=\textwidth]{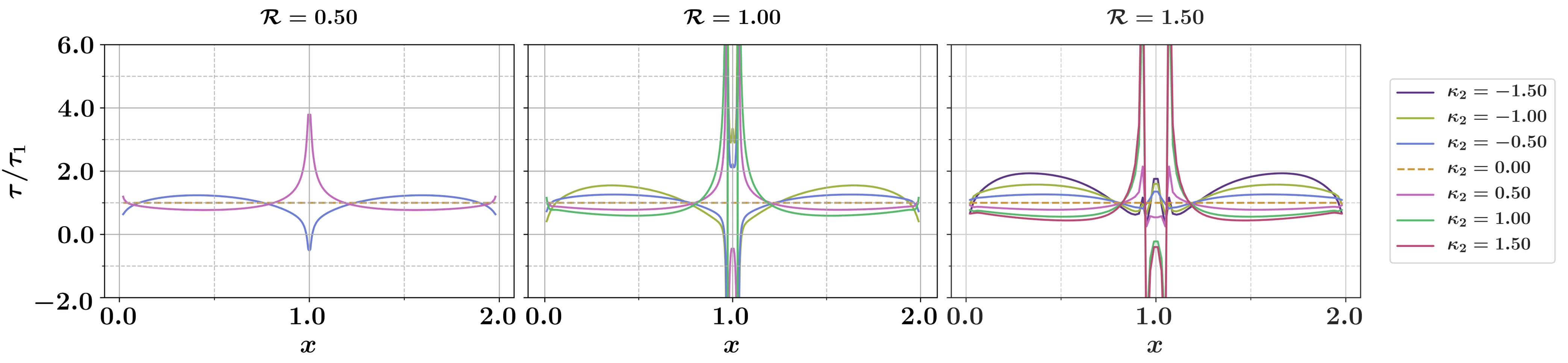}
\caption{Wall shear stress distribution within one texture element, point-wise normalized by the wall shear stress distribution of a triangular texture, $\tau_{_1}$, as a function of the spanwise direction, $x$, for textures defined with second order polynomials with height-to-half-spacing values of ${\cal R} = 0.50$, $1.00$, and $1.50$ and $\lambda/d = 0.26$.}
\label{WSS_second_order_T1}
\end{figure*}

\begin{figure*}[!ht]
\centering
\includegraphics[width=\textwidth]{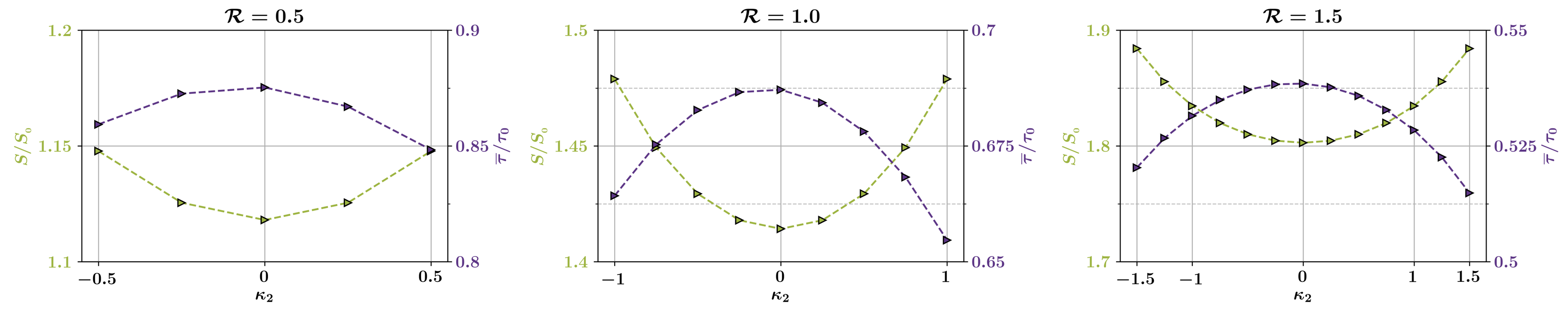}
\caption{Variations in the average wall shear stress, $\overline{\tau}$, normalized by the shear stress on a smooth rotor, $\tau_{_0}$ (right axis of the plots), experienced by textures defined by second order polynomials as a function of $\kappa_2$ with height-to-half-spacing ratios of ${\cal R} = 0.50$, $1.00$, and $1.50$ and $\lambda/d = 0.26$. The variations in the wetted surface area of the textures, normalized by the surface area of a smooth rotor, $S_0$, is also shown as a function of $\kappa_2$ (left axis of each plots).}
\label{WSS_2ndOrder_area}
\end{figure*}

Such variations in the iso-lines of the ($\eta-\xi$) coordinate system (Fig. \ref{meshes}) and the angular velocity distribution (Fig. \ref{v_theta_mesh}) as a function of $\kappa_2$ of these textures, directly results in a non-uniform velocity gradient and thus a non-uniform shear stress distribution on the wall as shown in Fig. \ref{WSS_second_order}(a) normalized by the shear stress experienced by a smooth rotor, $\tau_0$. The squeezed iso-lines at the peaks and the spaced out iso-lines in the valleys result in the shear stress at the peaks of all textures to be larger than the shear stress experienced by a smooth rotor $\tau_0$ (line $\tau/\tau_0 = 1.00$ in Fig. \ref{WSS_second_order}(a)), while in the troughs $\tau$ is lower than $\tau_0$ (similar to previous reports for triangular grooves \cite{choi1993direct, raayai2020geometry, vukoslavcevic1992viscous}).

The effect of the curvature of the texture, manifested in the $\kappa_2$ parameter, is visible in the local curvature of the iso-$\xi$ lines inside the grooves (the first iso-$\xi$ line is the texture boundary). In all cases, the iso-$\xi$ lines above the peaks have a negative local curvature (blue squares in Fig. \ref{meshes}) and as one moves toward the valleys the curvature slowly increases and turns positive (red circles in Fig. \ref{meshes}). As one moves from convex to concave textures, the extent of the segment of the iso-$\xi$ line with negative curvature slowly recedes and the portion of the iso-$\xi$ line with positive curvature increases. 

This variation in the local curvature of the iso-$\xi$ lines causes the distribution of the velocity gradient and the wall shear stress at the riblet-covered wall to span a range of below and above the shear stress on a smooth rotor, $\tau_{_0}$, and for ${\cal R} \geqslant 0.50$ the points of cross-over between $\tau>\tau_{_0}$ and $\tau<\tau_0$ takes place within $0<x \leqslant 0.503$ (Fig. \ref{WSS_second_order}(b)). In addition, the reduction in the portion of the iso-$\xi$ lines with negative curvature observed as $\kappa_2$ is increased directly results in the $x_{\rm cross-over}$ to be shifted closer to the peak of the grooves, and thus a larger segment of the textured wall exhibits shear stress values lower than ${\tau}_0$. This trend directly results in a decrease in the $G$ values for ${\cal R} \leqslant 1.00$ when $\kappa_2$ is increased. In addition, a point-wise comparison of the wall shear stress distribution of the textures with $\kappa_2 \neq 0$ and the wall shear stress of the triangular grooves, $\tau_{_1}$, (Fig. \ref{WSS_second_order_T1}) shows that a large segment of the wall shear of convex textures is placed higher than that of ${\tau}_{_1}$ and for concave ones, a large portion of each half of the texture has a wall shear stress lower than ${\tau}_1$. This ultimately results in convex textures having a larger torque compared to the triangular textures and for concave textures with ${\cal R} \leqslant 1.00$ to have lower torques than the triangular ones. However, for cases with ${\cal R} > 1.00$ the level of reduction in the local shear stress compared with $ {\tau}_1 $ observed in the ${\tau}<{\tau}_1$ portion of the texture for ${\kappa}_2>0$, is nearly similar for all the concave textures, and while the average wall shear stress has a decreasing trend, the total wetted surface area of the texture is increased (Fig. \ref{WSS_2ndOrder_area}), and thus no substantial difference with respect to the triangular textures is seen.

\subsection*{Inflection point}

\begin{figure*}[!ht]
    \centering
    \includegraphics[width = 1 \textwidth]{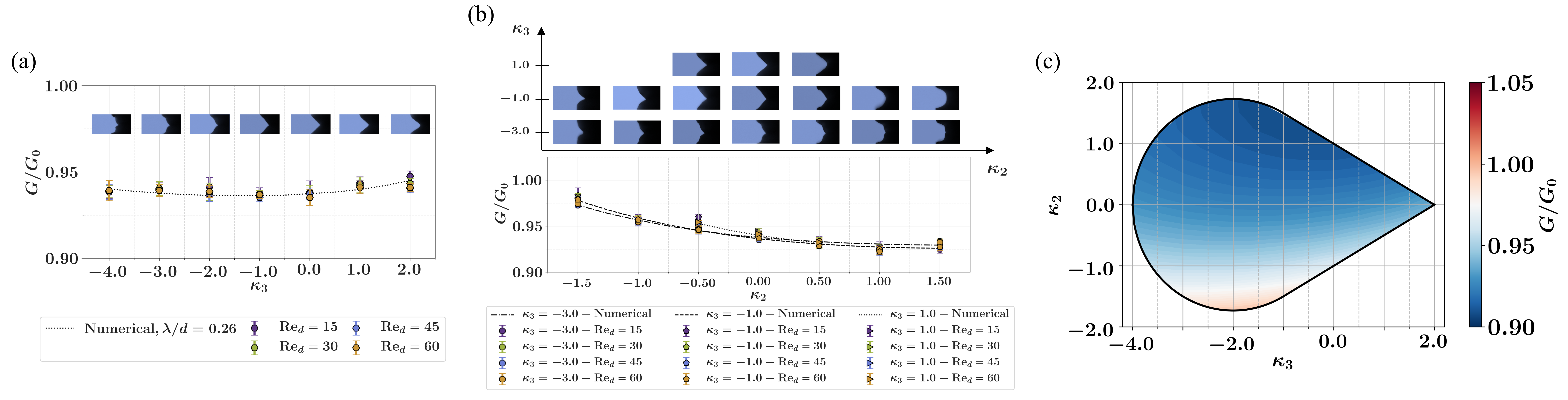}
    \caption{(a) Dimensionless torque experienced by textured rotors defined by third order polynomials, with ${\cal R} = 1.00$, $\kappa_2 = 0.00$, and $\lambda/d = 0.26$ as a function of $\kappa_3$. Torque is normalized by the torque experienced by the smooth rotor. Images of samples are presented above the corresponding points. (b) Dimensionless torque experienced by textured rotors with third order profiles, with ${\cal R} = 1.00$, $\lambda/d = 0.26$, and $\kappa_3 = -3.00$, $-1.00$, and $1.00$, as a function of $\kappa_2$. Torque is normalized by the torque experienced by the smooth rotor, $G_0$. Images of the samples are presented above the corresponding points. (c) Contours of dimensionless torque of textured rotors with a profile defined by third order polynomials with ${\cal R} =1.00$ as a function of $\kappa_3$, and $\kappa_2$, normalized by the torque of a smooth rotor, $G_0$, calculated numerically, for grooves with $\lambda/d = 0.26$.}
    \label{third_torques}
\end{figure*}

After concavity, the next level of complexity is the possibility of the profile $f$ to possess an inflection point and the simplest function for that is a third order polynomial of the form $a_3 x^3 + a_2 x^2 + a_1 x + a_0$. For a third order polynomial, assuming $d^3n_{\rm w}/dx^3 = 6 \kappa_3$, and an average concavity (the second derivative of the function is no longer constant) defined as  

\begin{equation}
    \dfrac{dn_{\rm w}}{dx} \bigg \vert _{x = 1^{-}} -  \dfrac{dn_{\rm w}}{dx} \bigg \vert _{x= 0^{+}} = 2 \kappa_2
\end{equation}

\noindent and height-to-half-spacing ratio of $\cal R$, we have: 

\begin{equation}
    n_{\rm w} = \kappa_3 x^3 + \left(- \dfrac{3}{2} \kappa_3 + \kappa_2 \right) x^2  + \left(\dfrac{\kappa_3}{2}-\kappa_2 - {\cal R} \right) x + n_0, \label{order_3}
\end{equation}

\noindent where at a constant $\cal R$, to keep the textures monotonic in $0\leqslant x \leqslant 1$, the limits of $\kappa_2$ and $\kappa_3$ are defined by

\begin{equation}
   - \left({\cal R}-\dfrac{{\kappa_3}}{2} \right) \leqslant \kappa_2 \leqslant  \left({\cal R}-\dfrac{{\kappa_3}}{2} \right)   
\end{equation}

\noindent for $-{\cal{R}} \leqslant \kappa_3 \leqslant 2{\cal{R}}$ and

\begin{equation}
- \sqrt{-3 \kappa_3 \left(\dfrac{\kappa_3}{4}+ {\cal R} \right)} \leqslant \kappa_2 \leqslant  \sqrt{-3 \kappa_3 \left(\dfrac{\kappa_3}{4}+ {\cal R} \right)}
\end{equation}

\noindent for $-4{\cal{R}} \leqslant \kappa_3 \leqslant -{\cal{R}}$ (cross-sectional images of 3D printed textures with $f$ defined by Eq. \eqref{order_3} as a function of $\kappa_2$ and $\kappa_3$ are presented in Fig. \ref{third_torques}(a) and \ref{third_torques}(b)). Note that if $\kappa_3 = 0$, then Eq. \eqref{order_3} returns back to Eq. \eqref{order_2}, and thus this definition encompasses all the textures defined by the lower order polynomials as well. While all third order polynomials have one inflection point in $\mathbb{R}$, it is not guaranteed that this point resides within the bounds of $0 \leqslant x \leqslant 1$ where profile $f$ is defined and for Eq. \eqref{order_3}, only profiles with $\vert \kappa_2/\kappa_3 \vert < 3/2$ have an inflection point in $0\leqslant x \leqslant 1$.

For textures defined by third order polynomials with ${\cal R} = 1$, and $\lambda/d = 0.26$, in the absence of mean concavity/convexity (i.e. $\kappa_2 = 0$), addition of the $\kappa_3$ term ($-4 \leqslant \kappa_3 \leqslant 2$) does not offer a significant difference in the torque compared to the triangular textures ($\kappa_2 = \kappa_3 = 0$ - Fig. \ref{third_torques}(a)) and $G$ in the CCF regime (independent of the Reynolds number) is only marginally affected by $\kappa_3$ as found by numerical simulations and experimental results, with the limiting ends ($\kappa_3 = -4$ and $\kappa_3 = 2$) offering slightly larger torque values.

 \begin{figure*}[!ht]
    \centering
    \includegraphics[width = 1 \textwidth]{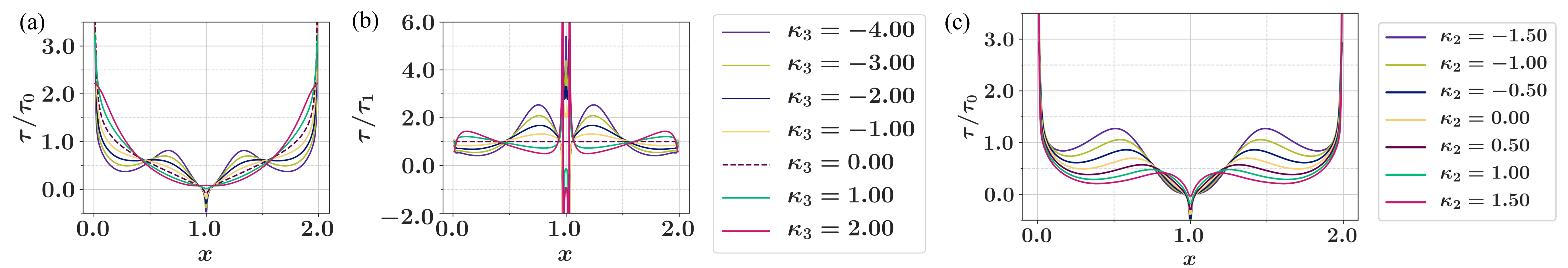}
    \caption{Distribution of wall shear stress for textures with ${\cal R} = 1.00$, $\lambda/2 = 0.26$, and $\kappa_2 = 0.00$, normalized by (a) the shear stress of smooth rotor, $\tau_0$, and (b) point-wise normalized by the shear stress of a triangular groove ($\kappa_2 = \kappa_3 = 0$), $\tau_1$, as a function of $\kappa_3$ and spanwise location, $x$. (Wall shear stress results for the triangular textures have been additionally marked with dashed lines.) (c) Distribution of wall shear stress  for textures defined by third order polynomials with ${\cal R} = 1.00$, $\lambda/2 = 0.26$, and $\kappa_3 = -3.00$ as a function of $\kappa_2$ and spanwise location, $x$. }
    \label{third_k3_all_2}
\end{figure*}

\begin{figure*}[!ht]
    \centering
    \includegraphics[width = 1 \textwidth]{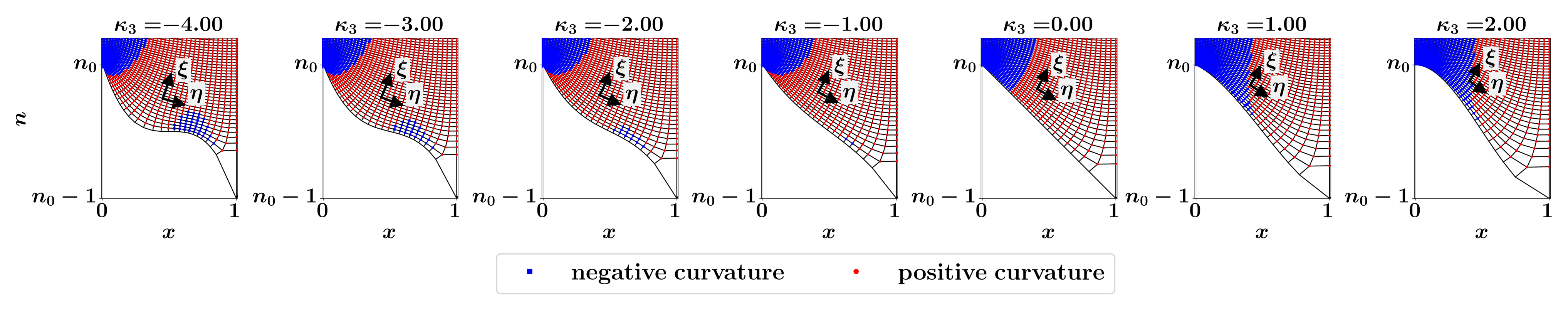}
    \caption{Iso-lines of curvilinear, orthogonal coordinate system $(\eta-\xi)$ inside grooves, defined by third order equation Eq. \eqref{order_3}, with ${\cal R} = 1$, $\lambda/d=0.26$, and $\kappa_2= 0$ as a function of $\kappa_3$. Red circles and blue squares denotes segments of the iso-$\xi$ lines with positive and negative curvature respectively.}
    \label{Third_order_mesh}
\end{figure*}

This negligible difference in the response of textures defined by this group of third order polynomials and that of a triangular profile ($\kappa_3 = 0.00$) is further assessed using the wall shear stress distribution at the textured wall normalized by the shear stress of a smooth rotor, $\tau_0$, and point-wise normalized by the shear stress of a triangular textures, $\tau_1$, both as a function of the spanwise direction (Figs. \ref{third_k3_all_2}(a) and \ref{third_k3_all_2}(b)). In all cases, the wall shear stress distribution oscillates about the wall shear stress of the triangular groove ($\kappa_3 = 0.00$, marked by dashed lines on Figs. \ref{third_k3_all_2}(a) and \ref{third_k3_all_2}(b)); for $\kappa_3<0$ the shear stress distribution twice crosses over the wall shear stress of the triangular textures where along $x$ within $0<x<1$, it is initially lower, then higher, and then lower than the wall shear stress of the triangular grooves. For $\kappa_3>0$, the cross-over happens three times where close to the peak, for a small portion, the wall shear stress is lower than the triangular case, and then moves to higher, then lower, and ultimately higher than the wall shear stress for the triangular textures. (Note that the point closest to $x=1$ that seems like a cross-over in $\tau/\tau_{_1}$ in Fig. \ref{third_k3_all_2}(b) is due to the $\tau_{_1}$ changing sign close to the trough and at this point $\tau/\tau_{_1} \rightarrow \infty$.) As a result of this oscillation for both positive and negative $\kappa_3$, the changes in the wall shear stress cancel each other out. In addition, while the average shear stress of each of these textures is lower than that of a triangular texture (Fig. \ref{WSS_third_area}(a)), the excess wetted surface area of the textures with $\kappa_3 \neq 0$ overcomes the decrease in the average shear stress. All these result in ultimately a negligible difference in the total frictional torque experienced by this family of textured walls as a function of $\kappa_3$.

While the cross-over of the wall shear stress distributions between $\tau>\tau_1.00$ and $\tau<\tau_1.00$ close to peaks and troughs are also observed in the profiles of textures defined by polynomials of second order (Figs. \ref{WSS_second_order} and \ref{WSS_second_order_T1}), the cross-over observed close to $x \approx 1/2$ is due to the presence of the inflection point at $x=1/2$. The effect of the inflection point is clear in the distribution and the curvature of the iso-$\xi$ lines (Fig. \ref{Third_order_mesh}) inside these grooves. For $\kappa_3<0$, the iso-$\xi$ lines are clustered close to each other at the peaks, and as one moves along $x$, they slowly become distant, until reaching close to the inflection point where the lines cluster together and then again grow apart close to the troughs. As a result of this, unlike the case of the profiles defined by second order polynomials, close to the texture wall for cases with $\kappa_3<0.00$, a portion of the iso-$\xi$ lines past the inflection point at $x=1/2$ has a negative curvature, which also corresponds to the region with $\tau/\tau_1>1.00$. 

Similarly, for the rest of the third order textures with ${\cal R} = 1$, $\lambda/d = 0.26$, $\kappa_3 \neq 0.00$ (Figs. \ref{third_torques}(b), \ref{third_k3_all_2}(c), and \ref{WSS_third_area}(b)), the trend in the measured torque in CCF regime is nearly insensitive to $\kappa_3$ and is primarily dependent on $\kappa_2$. Similar to the case of $\kappa_2=0.00$, the oscillation of the wall shear stress distribution of the cases with $\kappa_3 \neq 0.00$ around the wall shear stress distribution of the texture with the same $\kappa_2$ and $\kappa_3 =0.00$ results in nearly similar levels of torque among textures with constant $\kappa_2$ and different $\kappa_3$ terms. Also, gathering the results of the numerical simulations of all third order textures into a contour plot of normalized torque as a function of $\kappa_2$ and $\kappa_3$ (Fig. \ref{third_torques}(c)) shows that the changes are dominantly happening along the $\kappa_2$ axis and the changes along the $\kappa_3$ axis are minor. 

\begin{figure}[!ht]
    \centering
    \includegraphics[width = 0.35 \textwidth]{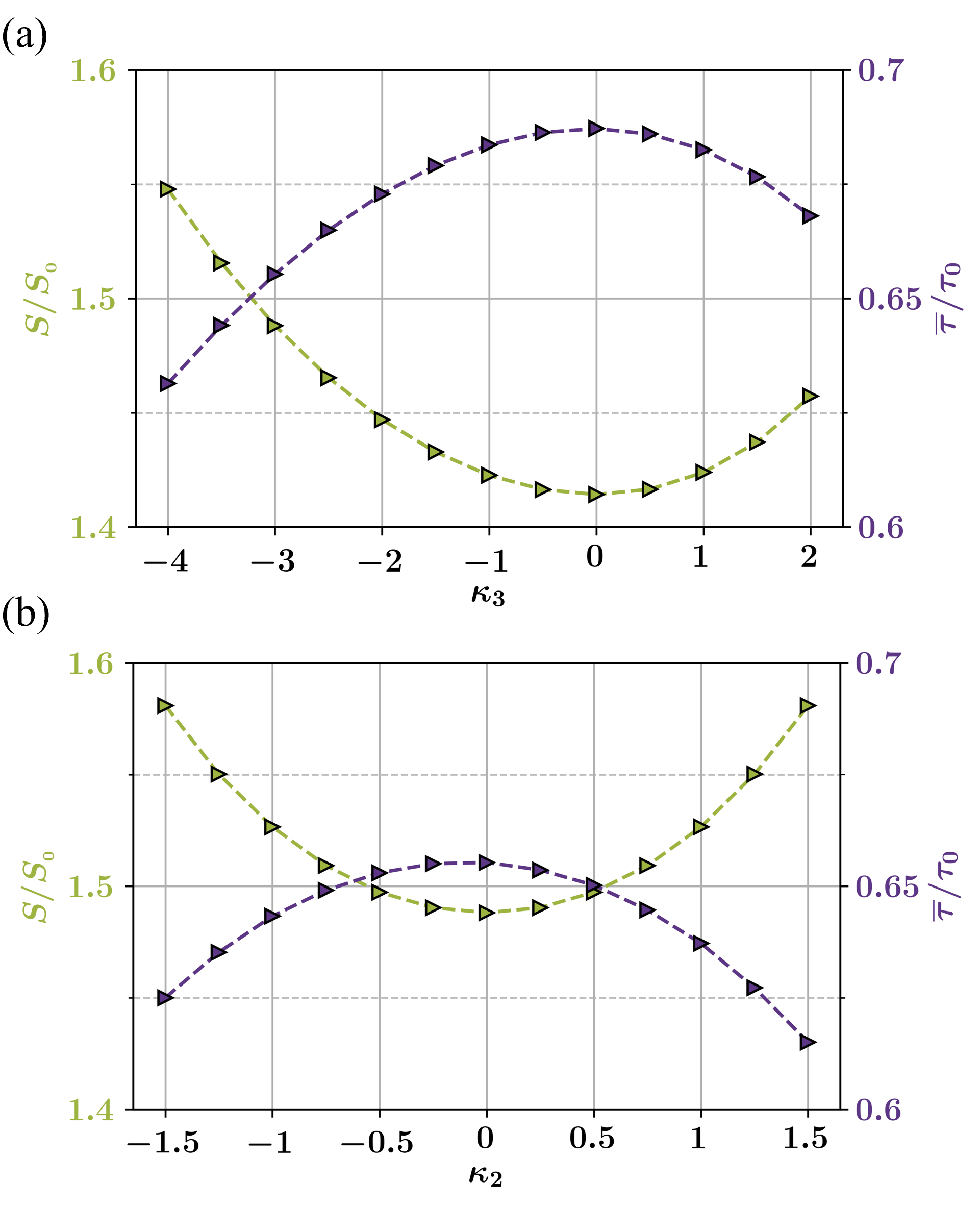}
    \caption{Variations in the average wall shear stress, $\overline{\tau}$, normalized by the shear stress on a smooth rotor, $\tau_{_0}$ (right axis of the plots), experienced by textures defined by third order polynomials with ${\cal R} = 1.00$, $\lambda/d = 0.26$, and (a) $\kappa_2=0.00$ as a function of $\kappa_3$, and (b) $\kappa_3 = -3.00$ as a function of $\kappa_2$ .The variations in the wetted surface area of the textures, normalized by the surface area of a smooth rotor, $S_0$, is also shown for both cases (left axis of each plots).}
    \label{WSS_third_area}
\end{figure}

\subsection*{A generalized mathematical framework for construction of symmetric and periodic 2D textures}

The complexity of a texture profile is then extended by considering a polynomial of $J$-th order. To generalize the definition, suppose we want to generate the profile of a texture, $f$, in the form of a function characterizing the local height of the texture in between each peak-trough ($P_i$-$T_i$) segment within $0 \leqslant x \leqslant 1$ (Fig. \ref{fig:peak_trough_schematic}(b)). Let the profile of the textures to be of the form of a polynomial  

\begin{equation}
    n_{\rm w} = \sum_{j=0}^{J} a_j x^j  \ \ \ \ 0 \leqslant x \leqslant 1 \label{polynom}
\end{equation}

\noindent where $J$ or the degree of polynomial is a measure of the geometric complexity of the texture. Coefficients $a_j$ can take any value, however, here, the domain of each of the coefficients is limited to ensure the all profiles pass though the chosen $P_i$ and $T_i$; the profiles are continuous everywhere; the derivatives of the profiles are continuous everywhere except at the peaks and troughs of the textures; chosen peaks and troughs of the function are the only global extrema of the function, and no local minima and/or maxima is allowed within $0<x<1$.

While any family of functions can be a valid choice for mathematically representing the shape of the profiles, polynomials not only have the ability to capture some of the most popular texture profiles (as shown earlier) but also allow for gradually increasing the complexity of the textures by changing the degree $J$; Using Eq. \eqref{polynom} with $J=0$ (the zeroth order), $n_w = n_0$ returns the smooth reference surface (i.e. case of ${\cal{R}} = 0$), and $J=1$ creates the V-groove (or triangular) profile with $a_0 = n_0$ and $a_1 = -{\cal R}$ (Eq. \eqref{order_1}). 

For the first order, height and spacing of the texture are sufficient to uniquely define the profiles and the coefficients $a_0$ and $a_1$. However, from the second order forward, there is no geometric guide dictating the boundary conditions to define the value of the coefficients and variations in values of $a_j$ ($j \geqslant 2$) results in more complicated texture profiles, while keeping the wavelength and height-to-half-spacing constant. At each $j$ level, the effect of $a_j x^j$ monomial can be decomposed into two portions, one part is entirely responsible for the effect of $j$th order and the other portion ensures that the profile passes through the given peaks and troughs.  
Let's assume that the portion of the $a_j$ coefficient responsible for the effect of the level of complexity, $j$, is represented by the parameter $\kappa_j$ (all $\kappa_j$ independent of each other), and is defined as the difference of the ($j-1$)th derivative of the profile at the (limits of) peak and trough normalized by $j!$. Thus for $2 \leqslant j \leqslant J$

\begin{equation}
    \dfrac{d^{(j-1)} n_{\rm w}}{dx^{(j-1)}}\bigg\rvert_{x=1^{-}} - \dfrac{d^{(j-1)} n_{\rm w}}{dx^{(j-1)}}\bigg\rvert_{x=0^{+}} = j! \ \kappa_{j} \label{k_2_n}
\end{equation}

\noindent and for $j=1$

\begin{equation}
    n_{\rm w}\bigg\rvert_{x=1} - n_{\rm w}\bigg\rvert_{x=0} = 1! \ \kappa_1.  \label{k_1}
\end{equation}

\noindent and $\kappa_0 = n_0$ is entirely dependent on the choice of the reference surface. As a result, coefficients $a_j$, can be decomposed into weighted sums of the $\kappa_{_l}$ where $l \geqslant j$, separating the effect of each order, 

\begin{equation}
    a_j = \sum_{l=j}^J  {\kappa}_{_l} m_{_{lj}} \label{coeff}
\end{equation}

\noindent and ultimately collecting all $m_{_{lj}}$ coefficients in matrix $\mathbf{\cal M}$ we can write $n_{\rm w}$ in the form

\begin{equation}
    n_{\rm w}(x) = \begin{bmatrix} \kappa_0 & \kappa_1 & \cdots & \kappa_{_J} \end{bmatrix} \mathbf{\cal{M}} \begin{bmatrix} x^0 \\ x^1 \\ \vdots \\ x^J \end{bmatrix}.
\end{equation}

\noindent Based on this definition, $\mathbf{\cal M}$ is a lower triangular matrix where $m_{_{00}} = m_{_{11}} = \dots = m_{_{JJ}} = 1$ and the rest of the coefficients of each column can be calculated using Eq. \eqref{k_2_n} and Eq. \eqref{k_1}. From the boundary conditions, for all textures we have $\kappa_1 = - {\cal R}$ and $\kappa_0 = {\cal R}$, and the rest of $\kappa_j$ can take any value within the physical ranges defined by the conditions described by the user.

The effect of this decomposition can be demonstrated using the two cases studied earlier; a second order profile of the form of Eq. \eqref{order_2}, is written as the sum of $\kappa_2 (x^2-x)$, and $- {\cal R} x+ n_0$. Rewritten this way, the second order profile is decomposed into the profile of a first order texture (Eq. \eqref{order_1}) and a term of second order entirely dependent on $\kappa_2$; the first monomial, $\kappa_2 x^2$ adds curvature to the profile and the second monomial $-\kappa_2 x$ ensures that the addition of the $\kappa_2 x^2$ does not violate the requirement of the profile passing through the designated peaks and troughs (Fig. \ref{decompose}(a)). Similarly, the third order profile of Eq. \eqref{order_3} can be written as the sum of $\kappa_3 \left( x^3 -{3}/{2} x^2 +{1}/{2} x \right)$ (a third order part identified by the $\kappa_3$) and Eq. \eqref{order_2} (Fig. \ref{decompose}(b)). Thus, this decomposition, separates the effect of each level of complexity while also including the effect of lower order terms within its definition. This results in textures defined by lower order polynomials to be subsets of those defined by higher orders and allowing for easier comparison across shapes (i.e. setting $\kappa_{_J} = 0$ returns the family of textures defined by order $J-1$). Thus, the decomposition of the coefficients of the polynomials using this definition of the $\kappa_{_j}$ allows for successive expansion of the texture profiles one order at a time toward more complicated shapes, while keeping the lower order ones among the possible variations.

\begin{figure}[!ht]
    \centering
    \includegraphics[width = 0.45 \textwidth]{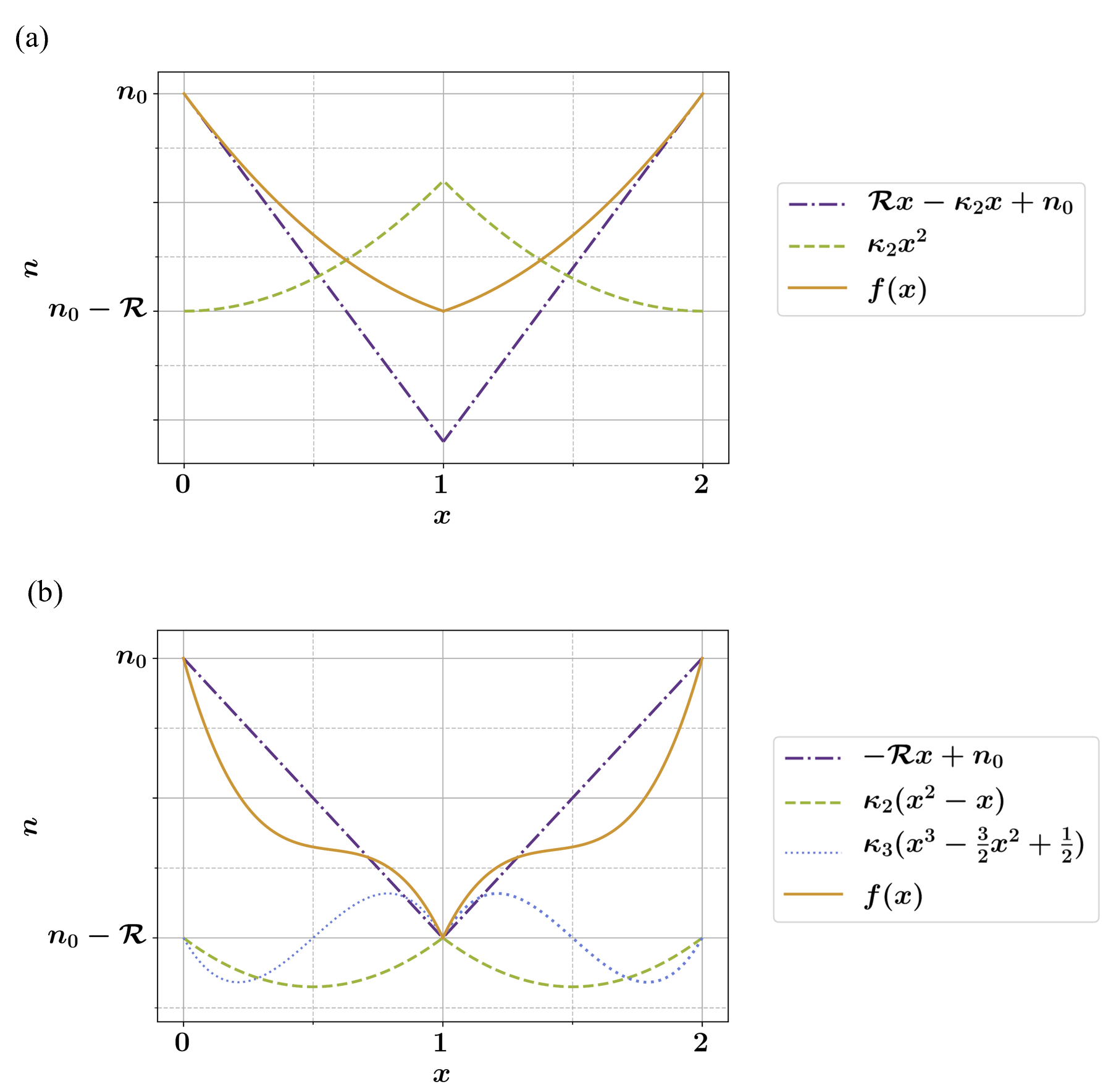}
    \caption{(a) Profile of an example texture, $f(x)$, defined by a second order polynomial decomposed into terms identified by $x^0$ and $x$, and $x^2$. Note that with $\kappa_2 x^2$ term alone, the profile cannot maintain the location of the peak and trough of the texture, and the $-\kappa_2 x$ monomial is required to satisfy this boundary condition. (b) Profile of an example texture, $f(x)$, defined by a third order polynomial decomposed into terms identified by $\kappa_0 = n_0$ and $\kappa_1 = -{\cal R}$, $\kappa_2$, and $\kappa_3$.}
\label{decompose}
\end{figure}

\begin{figure*}[!ht]
\centering
\includegraphics[width=\textwidth]{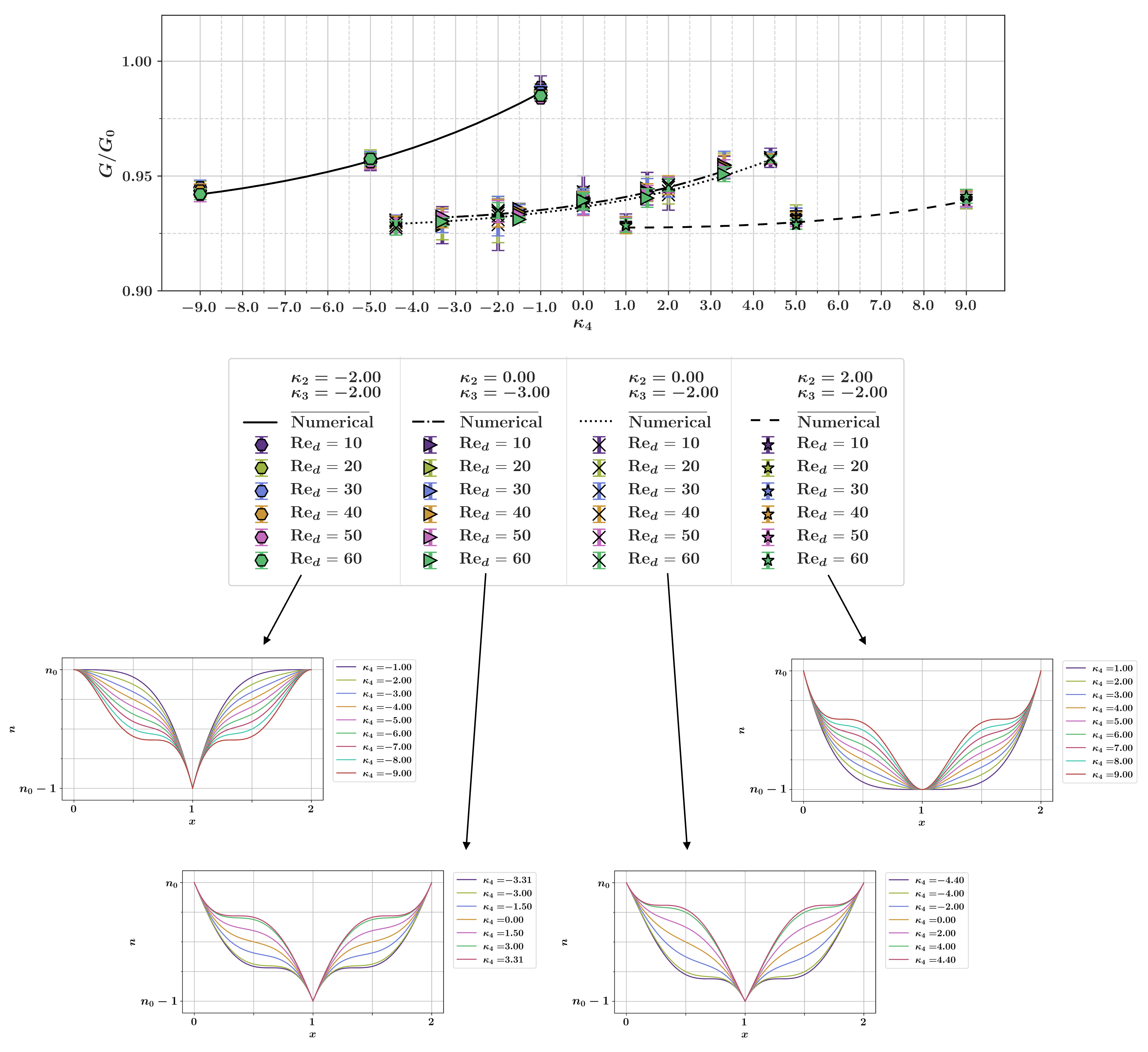}
\caption{Torque experienced by textured rotors with profiles defined by fourth order polynomials (Eq. \eqref{order_4}) as a function of $\kappa_4$, normalized by the torque of a smooth rotor, $G_0$, with ${\cal R} = 1.00$, $\lambda/d = 0.26$, and various values of $\kappa_2$ and $\kappa_3$. Variations in the shape of the profiles as a function of all the $\kappa$ parameters of each group are also shown below the plot.}
\label{4th_torque}
\end{figure*}

\begin{figure*}[!ht]
\centering
\includegraphics[width=\textwidth]{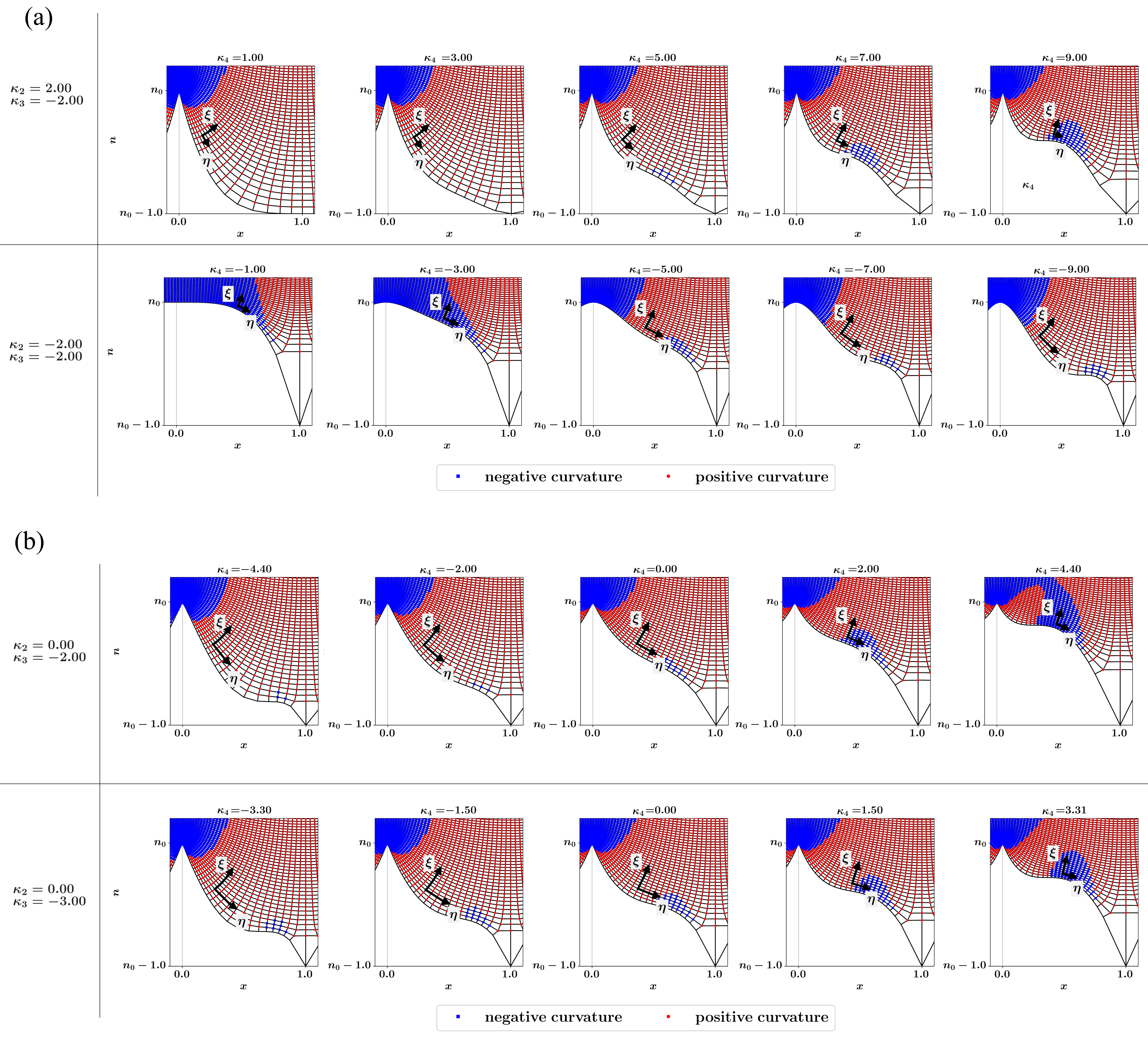}
\caption{Curvilinear orthogonal coordinates $(\eta-\xi)$ for textures defined by fourth order polynomials with ${\cal R} = 1.00$ (a) $\kappa_2 = \pm 2.00$, and $\kappa_3 = -2.00$, and (b) $\kappa_2 = 0.00$, $\kappa_3 = -2.00$ and $-3.00$  as a function of $\kappa_4$ values within their allowed limits.}
\label{4th_iso_lines}
\end{figure*}

Polynomials of order higher than 3, offer a wide range of possibilities to design texture profiles, however, the increase in the complexity does not offer a substantial difference in the hydrodynamic response of these textures in the CCF when compared with the second order textures. Results of experiments and simulations with textures defined by fourth order polynomials of the form

\begin{equation}
        n_{\rm w}  = \kappa_4 (x^4 - 2 x^3 + x^2) + \kappa_3 \left(x^3-\dfrac{3}{2}x^2+\dfrac{1}{2}x \right) +  \kappa_2 (x^2-x)  - {\cal R}  x + n_0 
        \label{order_4}
\end{equation}

\noindent presented in Fig. \ref{4th_torque}, show that variations in the torque measured with such textures are mainly dictated by $\kappa_2$, and as $\kappa_2$ value is decreased, the effect of $\kappa_4$ becomes important only in offering a wider variation in the torque, but the variation in the values of torque recorded using textures with fourth order polynomials are within the same range as the variations seen with the second order profiles. In textures with a positive average concavity ($\kappa_2>0$) and a constant $\kappa_3$, increase in $\kappa_4$ only offers a slight increase in the torque, and analysis of the iso-$\xi$ lines of this sub-family of textures confirms the expansion of a region with negative curvature around $x=1/2$ which results in the slight increase in the torque (Fig. \ref{4th_iso_lines}(a)). In the absence of average concavity ($\kappa_2 = 0$), the response of the textures are entirely dependent on the $\kappa_4$ parameter and independent of $\kappa_3$, and as $\kappa_4$ is increased, the portion of the region close to the wall where the iso-$\xi$ lines have a negative curvature increases. As a result, within this sub-family of textures, increase in $\kappa_4$ results in an increase in the torque (Figs. \ref{4th_torque} and \ref{4th_iso_lines}(b)). As one moves toward $\kappa_2<0$ cases, the effect of $\kappa_4$ becomes more pronounced allowing for a wider range of torque responses compared to the cases with $\kappa_2 \geqslant 0$ (Figs. \ref{4th_torque} and \ref{4th_iso_lines}(a)). Overall, in all cases, at constant $\kappa_2$ and $\kappa_3$, textures with the lowest $\kappa_4$ offer the lowest torque of the sub-family, and increasing $\kappa_4$ results in an increase in the portion of iso-$\xi$ close to the wall with negative curvature which directly results in an increase in the total frictional torque within each sub-family considered (Fig. \ref{4th_iso_lines}).

\subsection*{Families of extreme concave/convex textures}

The definition of textures based on polynomials of higher orders allows us to consider families of textures that constitute the two most extreme cases of each order $J$: the case of textures that are either extremely concave 
defined by $n_{\rm w} = {\cal R} (1-x)^J + (n_0-{\cal R})$ or extremely convex given by $n_{\rm w} = -{\cal R} x^J + n_0$. (Figs. \ref{concave_convex}(a) and \ref{concave_convex}(b) -  $\kappa_j$ coefficients of these textures up to $J = 10$, for ${\cal R} = 1$, are listed in Tbls. \ref{coeff_concave} and \ref{coeff_convex}.) In the extreme concave case as $J$ is increased toward infinity the shape of the texture resembles an idealized form of rectangular (razor blade) textures \cite{el2007drag,Bechert2000} and for the extreme convex ones as $J$ is increased toward infinity, the profile asymptotically reaches back to the case of a smooth surface. In the concave group, as the order $J$ of the polynomial is increased, the portion of the texture in the vicinity of the trough is smoothed more and more (Fig. \ref{concave_convex}(a)), and in the convex group, increase in the order of the polynomial results in a wider smooth area at the peak of the textures (Fig. \ref{concave_convex}(b)). While the increase in the polynomial order of these textures requires additional $\kappa_j$ terms to define the profile, the increase in the concavity/convexity of these profiles is directly manifest in the increase in the magnitude of the  $\kappa_2$ term of each profile (Tbls. \ref{coeff_concave} and \ref{coeff_convex}).

\begin{figure*}[!ht]
    \centering
    \includegraphics[width = 1 \textwidth]{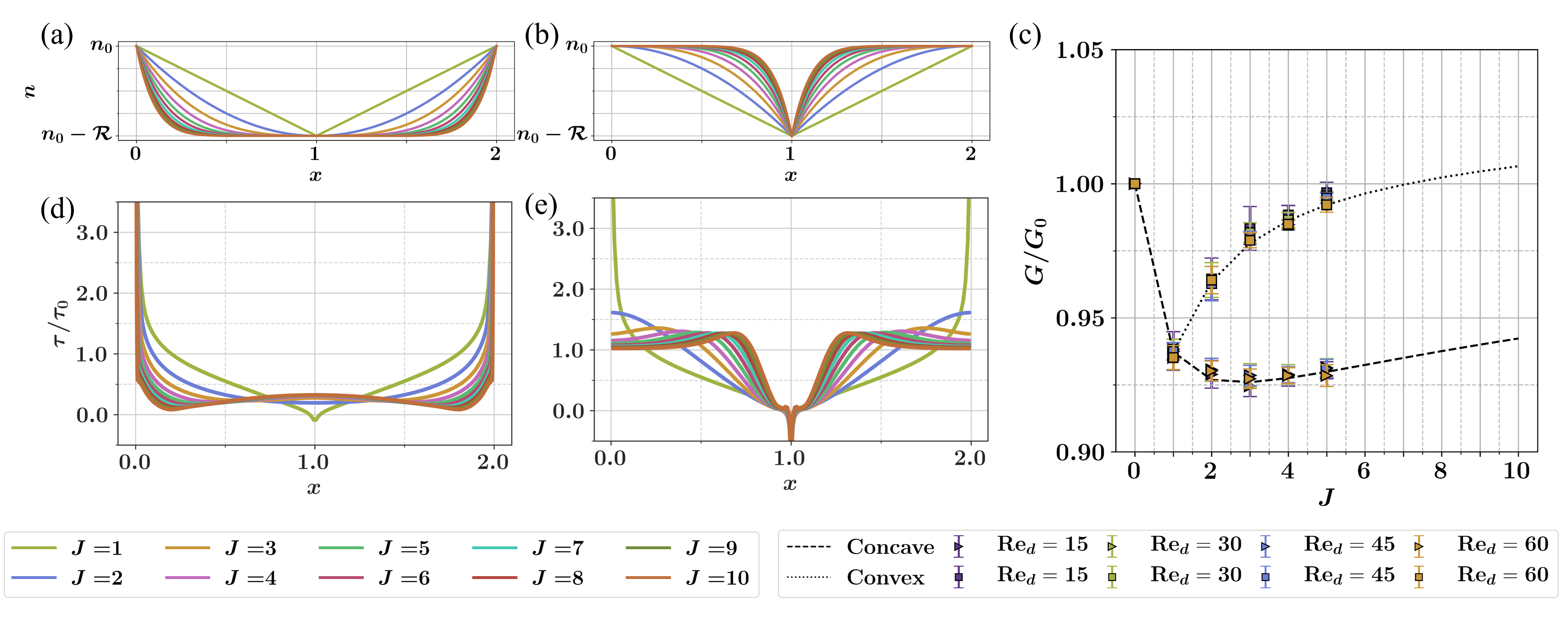}
    \caption{Schematic of families of (a) extreme concave and (b) extreme convex textures of orders 2-10 with arbitrary height-to-half-spacing of $\cal R$. Triangular groove ($J=1$) is shown for reference. (c) Torque experienced by extreme concave (triangular symbol) and extreme convex (rectangular symbol) textures of ${\cal R} = 1$, and $\lambda/d = 0.26$, normalized by the torque of a smooth surface. Torque values of a smooth surface and triangular texture ($J = 0$ and $J = 1$) are shown as reference. Results of the direct numerical simulation of concave and convex textures up to $J = 10$ are shown with dashed and dotted lines respectively. Distribution of the wall shear stress normalized by the wall shear stress of a smooth surface as a function of the spanwise direction for (d) extreme concave and (e) extreme convex textures with ${\cal R} = 1.00$, and $\lambda/d = 0.26$. The results for $J=1$ is presented for reference.
    }
    \label{concave_convex}
\end{figure*}

In the case of extreme concave textures ($J \geqslant 2)$, with ${\cal R} = 1.00$, and $\lambda/d = 0.26$, textures of order 2 and 3 have a nearly similar torque response (Fig. \ref{concave_convex}(c)) which is less than the case of the triangular textures and then beyond that as the order $J$ is increased, the total frictional torque is also increased, and after $J = 7$ it passes to higher than the torque experienced by a triangular texture. On the extreme convex side, for $J \geqslant 2$ as $J$ is increased (Fig. \ref{concave_convex}(c)), the torque is also increased and is always higher than torque of a triangular texture and by $J=7$ it crosses the torque of a smooth surface. 

\begin{table*}[!ht]\centering
\caption{${\kappa}_j$ values for extreme concave textures for polynomial orders 0-10 for ${{\cal R} = 1}$}
\label{coeff_concave}
\begin{tabular}{lrrrrrrrrrrr}
Polynomial Order & $\kappa_0$ & $\kappa_1 = -{\cal R}$ & $\kappa_2$ & $\kappa_3$ & $\kappa_4$, & $\kappa_5$, & $\kappa_6$, & $\kappa_7$ & $\kappa_8$, & $\kappa_9$, & $\kappa_{10}$ \\
\hline
$J=0$ & $n_0$-1 \\
$J=1$ & $n_0$-1 & -1 \\
$J=2$ & $n_0$-1 & -1 & 1 \\
$J=3$ & $n_0$-1 & -1 & 3/2 & -1 \\
$J=4$ & $n_0$-1 & -1 & 2 & -2 & 1\\
$J=5$ & $n_0$-1 & -1 & 5/2 & -10/3 & 5/2 & -1 \\
$J=6$ & $n_0$-1 & -1 & 3 & -5 & 5 & -3 & 1 \\
$J=7$ & $n_0$-1 & -1 & 7/2 & -7 & 35/4 & -7 & 7/2 & -1 \\
$J=8$ & $n_0$-1 & -1 & 4 & -28/3 & 14 & -14 & 28/3 & -4 & 1\\
$J=9$ & $n_0$-1 & -1 & 9/2 & -12 & 21 & -126/5 & 21 & -21 & 4.5 & -1\\
$J=10$ & $n_0$-1 & -1 & 5 & -15 & 30 & -42 & 42 & -30 & 15 & -5 & 1\\
\hline
\end{tabular}

\caption{${\kappa}_j$ values for extreme convex textures for polynomial orders 0-10 for ${{\cal R} = 1}$}
\label{coeff_convex}
\begin{tabular}{lrrrrrrrrrrr}
Polynomial Order & $\kappa_0$ & $\kappa_1 = -{\cal R}$ & $\kappa_2$ & $\kappa_3$ & $\kappa_4$, & $\kappa_5$, & $\kappa_6$, & $\kappa_7$ & $\kappa_8$, & $\kappa_9$, & $\kappa_{10}$\\
\hline
$J=0$ & $n_0$-1 \\
$J=1$ & $n_0$-1 & -1 \\
$J=2$ & $n_0$-1 & -1 & 1 \\
$J=3$ & $n_0$-1 & -1 & -3/2 & -1 \\
$J=4$ & $n_0$-1 & -1 & -2 & -2 & -1\\
$J=5$ & $n_0$-1 & -1 & -5/2 & -10/3 & -5/2 & -1 \\
$J=6$ & $n_0$-1 & -1 & -3 & -5 & -5 & -3 & -1 \\
$J=7$ & $n_0$-1 & -1 & -7/2 & -7 & -35/4 & -7 & -7/2 & -1 \\
$J=8$ & $n_0$-1 & -1 & -4 & -28/3 & -14 & -14 & -28/3 & -4 & -1\\
$J=9$ & $n_0$-1 & -1 & -9/2 & -12 & -21 & -126/5 & -21 & -21 & -4.5 & -1\\
$J=10$ & $n_0$-1 & -1 & -5 & -15 & -30 & -42 & -42 & -30 & -15 & -5 & -1\\
\hline
\end{tabular}
\end{table*}

For the family of extreme concave textures ($J \geqslant 2$), at the trough, the slope of the profiles come to zero and as $J$ is increased a larger segment of the profile in the vicinity of $x=1$ has a near-zero slope. Reverse of this trend applies to the peaks of extreme convex profiles where near the peaks the segment of the profile with near-zero slope is increased as $J$ is increased. This directly affects the wall shear stress distribution (Fig. \ref{concave_convex}(d)) where in the extreme concave cases for $J \geqslant 2$, the local shear stress at the trough of the grooves consistently increases as $J$ is increased and only for $J=2$, the local shear stress at the trough is minimum within $0 \leqslant x \leqslant 2$ and the minimum local wall shear stress for the rest of the profiles is pushed away from the trough and closer to the peaks. Thus, as order $J$ is increased the average shear stress within the texture profile $\overline{\tau_{_J}}$ takes a decreasing trend (Fig. \ref{Avg_concave_convex}).  However, as $J$ is increased, the increase in the wetted surface area results in the the total frictional torque to be the lowest for orders 2 and 3, and then to increases as $J$ is increased. On the extreme convex side, as the order is increased the local wall shear stress at the peak is slowly reduced and asymptotically reaches the value of the wall shear stress of a smooth surface (Fig. \ref{concave_convex}(e)). However, due to the excess wetted surface area compared to the smooth surface, for $J>7$ the total frictional torque is slightly larger than that of a smooth surface. With average shear stress $\overline{\tau_{_J}}$ of the convex textures being always larger than that of the concave textures (Fig. \ref{Avg_concave_convex}), even with the decreasing trend (as $J$ is increased), with the increase in the wetted surface area, torque experienced by convex textures takes a monotonically increasing trend as the order is increased ($J>1$).

\begin{figure}[!ht]
    \centering
    \includegraphics[width = 0.45 \textwidth]{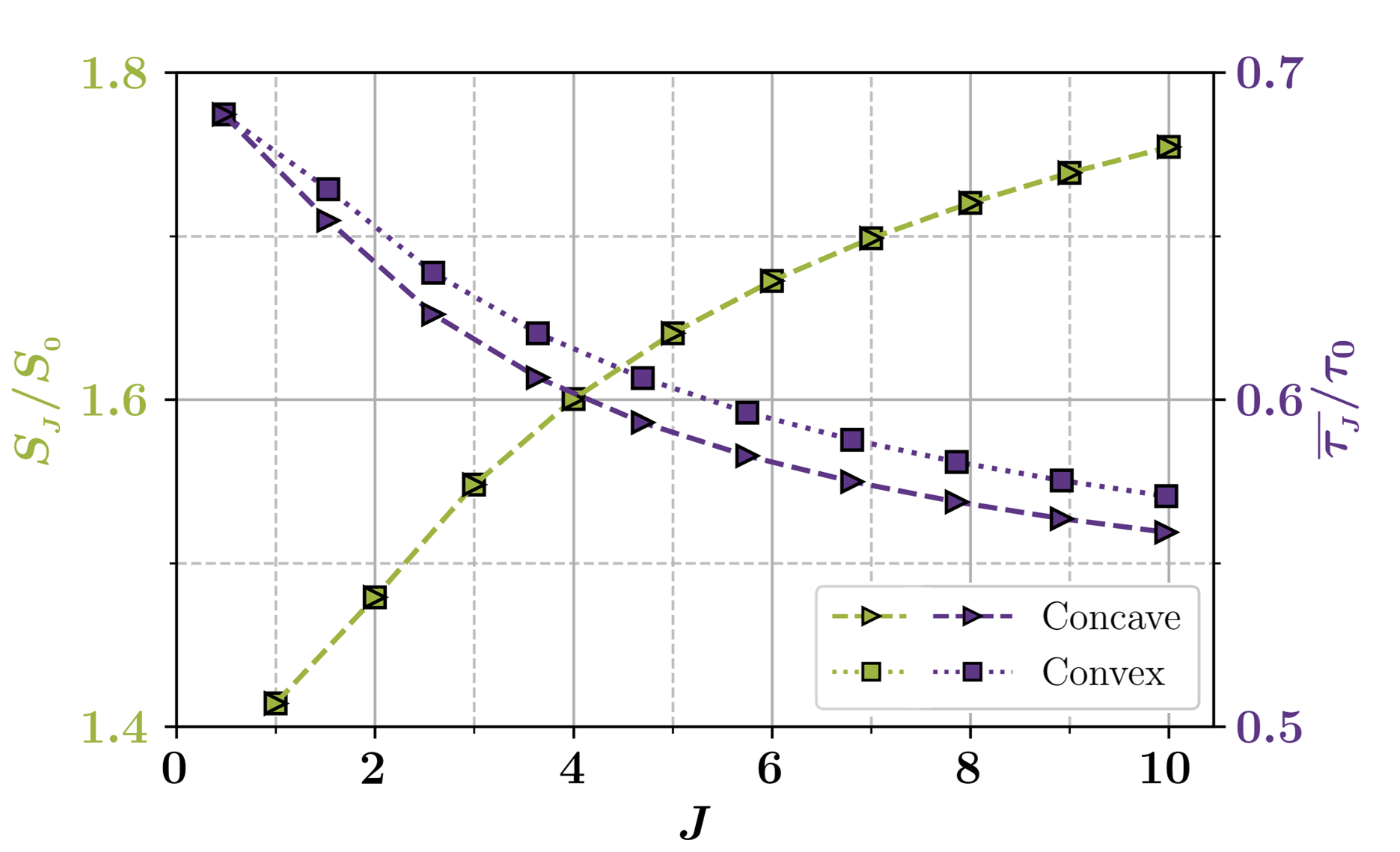}
    \caption{Total wetted surface area, $S_{_J}$ (left axis), and the average shear stress, $\overline{\tau_{_J}}$ (right axis) of extreme concave (triangular symbols) and convex (square symbols) textures as a function of their order $J$ for $J \geqslant 1$. Wetted surface area is normalized by the wetted surface area of a smooth reference rotor ($S_0$), and the average shear stress is normalized by the shear stress on a smooth rotor, $\tau_0$. It is clear that for $J=0$, $S_{_J}/S_0 = 1$ and $\overline{\tau_{_J}}/\tau_0 = 1$ (not shown).}
    \label{Avg_concave_convex}
\end{figure}

Ultimately, the comparison of the two extreme concave and convex textures with each other and with the smooth wall and the triangular texture shows that having a near smooth peak is more detrimental to the frictional torque experienced by a texture than having a near smooth trough, and while not the most ideal case, the extreme-concave textures can maintain a torque lower or equal to the level of the triangular textures for $J<8$ while extreme convex textures can barely maintain the reduction and are not well suited for the purpose of drag reduction in the CCF regime. This can also be a reason for natural textures on shark denticles \cite{wen_biomimetic_2014, domel2018hydrodynamic}, and peristome of the pitcher plants \cite{bohn_insect_2004,gorb_structure_2004,gorb_structure_nodate, bauer_insect-trapping_2009} having a concave form resembling profiles of $J=2$  or $J=3$, as opposed to triangular or convex shapes.

\section*{Discussion \& conclusion}

The decomposition of a texture profile into the effects of the $\kappa_j$ parameters, allows for a more comprehensive understanding of the effect of the geometry of the boundary on the physical response of the surface. Here, with a focus on the hydrodynamic response of the surface in laminar flows, it is clear that after the height-to-half-spacing, the $\kappa_2$ parameter, defining the average level of concavity of a texture plays the most important role in controlling the total frictional response of the surfaces. Textures defined by second order polynomials expand the range of frictional responses compared to the conventional V-grooves, but moving to third and fourth order polynomials does not offer a visible improvement compared to those defined by second order polynomials.

The proposed decomposition can be employed in reverse for any known texture of given profile $n_{w} = g(x)$ by fitting a polynomial of degree $J$ using a spline interpolation. If the coefficients of the $x^j$ in the spline are $b_j$, then the values of $\kappa_j$ terms for this profile can be found by solving the inverse of the linear system of equations in Eq. \eqref{coeff}, in the form ${\boldsymbol \kappa} = \mathbf{\cal {M}}^{-T} \mathbf{b}$ where $\boldsymbol \kappa$ and $\mathbf{b}$ are the vectors consisting of $\kappa_j$ terms and $b_j$ terms respectively.

It should be emphasized that the shape of polynomials at each degree enables us to create textures with profiles of various levels of complexity and assign quantitative labels to them. As a result of this framework, we are able to move away from qualitative description of texture profiles and track the effect of the geometry not only in terms of the key dimensions of the texture, but also in terms of the dimensionless $\boldsymbol \kappa$ vector that defines all the fine details of the shape of the profiles. 

While this work focuses on the hydrodynamic response of surfaces, this framework can easily be expanded to other physical problems where textured surfaces play a role and the effect of each of $\kappa_j$ parameters can be assessed based on the objectives of the problem of interest. }

\section*{Declaration of Competing Interest}
Author declares no conflict of interest. 

\section*{Acknowledgements}
This work was supported by the Rowland Fellows program at Harvard University.



 \bibliographystyle{elsarticle-num} 
 \bibliography{refs.bib}





\end{document}